\documentclass{aa}
\usepackage{fixltx2e}
\usepackage{graphicx}
\usepackage{txfonts}
\usepackage{rotating}

\newcommand{\Hii}{\ion{H}{ii} }
\begin{document}

\title{ALMA reveals the magnetic field evolution \\in the high-mass star forming complex G9.62+0.19}
\titlerunning{Magnetic field in G9.62}
\author{D. Dall'Olio\inst{1}
         \and
          W. H. T. Vlemmings\inst{1}
         \and
          M. V. Persson\inst{1}
          \and
          F. O. Alves\inst{2}
          \and
          H. Beuther\inst{3}
          \and
          J. M. Girart\inst{4,5}
          \and \\
          G. Surcis\inst{6}
          \and 
          J. M. Torrelles\inst{4,5}
          \and
          H. J. Van Langevelde\inst{7,8}
            }

\institute{Department of Space, Earth and Environment, Chalmers University of Technology,
            Onsala Space Observatory,\\ Observatoriev\"agen 90, 43992 Onsala, Sweden;
             \email{daria.dallolio@chalmers.se}
             \and
             Max-Planck-Institut f{\"u}r extraterrestrische Physik, Giessenbachstr. 1, D-85748 Garching, Germany
              \and
              Max-Planck-Institute for Astronomy, K{\"o}nigstuhl 17, 69117 Heidelberg, Germany
              \and
              Institut de Ci{\`e}ncies de l’Espai (ICE, CSIC), Can Magrans, s/n, E-08193 Cerdanyola del Vall{\`e}s, Catalonia 
              \and
              Institut d’Estudis Espacials de Catalunya (IEEC), E-08034, Barcelona, Catalonia
              \and
              INAF--Osservatorio Astronomico di Cagliari, Via della Scienza 5, 09047 Selargius, Italy
              \and
              Joint Institute for VLBI ERIC, Oude Hoogeveensedijk 4, 7991 PD Dwingeloo, The Netherlands
              \and
               Sterrewacht Leiden, Leiden University, P.O. Box 9513, NL-2300 RA Leiden, The Netherlands
               }

           \date{Received giorno mese anno; accepted giorno mese
             anno}

           \abstract {The role of magnetic fields during the formation
             of high-mass stars is not yet fully understood, and the
             processes related to the early fragmentation and collapse
             are as yet largely unexplored. The high-mass star forming
             region G9.62+0.19 is a well known source, presenting
             several cores at different evolutionary
             stages.}
           {We seek to investigate the magnetic field
             properties at the initial stages of massive star
             formation. We aim to determine the magnetic field
             morphology and strength in the high-mass star forming
             region G9.62+0.19 to investigate its relation to the
             evolutionary sequence of the cores.}
           {We made use of
             Atacama Large Millimeter Array (ALMA) observations in
             full polarisation mode at 1 mm wavelength (Band 7) and we analysed
             the polarised dust emission. We estimated the magnetic field
             strength via the Davis-Chandrasekhar-Fermi and structure function methods.}
           {We resolve several protostellar cores embedded in a bright
             and dusty filamentary structure. The polarised emission is clearly
             detected in six regions: two in the northern field
             and four in the southern field.
             Moreover the magnetic field is orientated along the filament and appears
             perpendicular to the direction of the outflows. The polarisation
             vectors present ordered
             patterns and the cores showing polarised emission are less fragmented.
             We suggest an evolutionary sequence
             of the magnetic field, and the less evolved hot core exhibits a
             stronger magnetic field than the more evolved hot core.
             An average magnetic field strength of the order of 11 mG was derived,
             from which we obtain
             a low turbulent-to-magnetic energy ratio,
             indicating that turbulence does not significantly
             contribute to the stability of the clump. We report a detection of
             linear polarisation from thermal line emission, probably from methanol or
             carbon dioxide, and we tentatively compared
             linear polarisation vectors from our observations with previous
             linearly polarised OH masers observations.  We also compute the spectral
             index, column density, and mass for
             some of the cores.}
             {The high magnetic
             field strength and smooth polarised emission indicate
             that the magnetic field could play an important role in the
             fragmentation and the collapse process in the star forming region G9.62+019 and that
             the evolution of the cores can be magnetically regulated. One core shows a very
             peculiar pattern in the polarisation vectors, which can indicate a compressed magnetic field.
             On average, the magnetic field derived by the linear
             polarised emission from dust, thermal lines, and masers is pointing in
             the same direction and has consistent strength.}
    
   \keywords{magnetic field --
                stars: formation -- stars: massive --
                dust -- polarisation}

   \maketitle

\section{Introduction}

High-mass stars play a fundamental role in the evolution of the
universe, but despite their importance, our comprehension of the
physical mechanisms occurring in their formation is still incomplete.
As a consequence, the high-mass star formation theory has been a topic
of great debate in recent decades, and several models have been
proposed such as the core accretion model (e.g.\,
\citealt{McKeeTan2002}, \citealt{Banerjee2007}) and the competitive
accretion model (e.g.\, \citealt{Bonnell2006}). In particular, one of
the subjects under discussion is the role played by the magnetic field
at the first stages of massive star formation.  While some theoretical
and simulation works showed that magnetic fields can significantly
influence each stage of massive star formation (e.g.\,
\citealt{Mouschovias1979, Mouschovias2006, Commercon2011, Tan2013,
  Tassis2014, Klassen2017}), other authors consider the effect of
turbulence and gravity more important (e.g.\, \citealt{Padoan2002,
  Klessen2005, Vazquez-Semadeni2011, Wang2014}).

Recently, some theoretical simulations of molecular cloud collision
show that massive star formation can be enhanced by the strong
magnetic fields generated in the shocked region \citep{Inoue2018,
  Inoue2013, Vaidya2013}. Moreover, other papers indicate that
magnetic fields affect fragmentation, slow down cloud collapse,
influence accretion, and drive feedback phenomena such as collimated
outflows, which are important in removing excess angular momentum
\citep{McKee2007, Myers2013, Peters2014, Susa2015, Matsushita2018}.
These works indicate that turbulence can play a major role during the
initial collapse in the regions where massive stars form, while
magnetic fields are likely to be important close to the protostar, as
evidenced by the detection of a pinched magnetic field in a few
massive protostars (e.g.\,\citealt{Girart2009, Qin2010}).
Furthermore, magnetic fields could also be required to stabilise discs
and allow for the large accretion rate necessary during massive star
formation \citep{Johansen2008,Stepanovs2014}.  In addition, some
authors suggest that non-ideal magnetohydrodynamic (MHD) effects such
as ambipolar diffusion or Ohmic dissipation can explain some observed
configurations such as accretion discs and outflows
\citep{Zhao2016,Machida2014,Seifried2012}.

However it is still unclear which process dominates at which
evolutionary stage, and more investigations are needed to overcome
this uncertainty and to estimate the magnitude and morphology of
the initial magnetic field.

A possible way to study the magnetic field parameters is through the
analysis of dust polarisation observations.  Interstellar dust
produces thermal emission and extinction of light from a background
star. The magnetic field morphology in the interstellar medium (ISM)
can be probed by observing the linear polarisation of this
radiation. This polarisation requires the alignment of irregularly
shaped grains, as described by for example\ \citet{Cudlip1982,
  Hildebrand1988, Draine1996, Draine1997, Lazarian2000, Cho2005,
  Lazarian2007, Lazarian2008, Hoang2008}, and
\citet{Andersson2015}. The first successful observation of linearly
polarised dust emission towards a molecular cloud was made by
\citet{Hildebrand1984}, and since then many other studies were
presented (e.g.\,\citealt{Crutcher2012}, and references
therein). Interferometric images of the polarised emission of dust can
show the magnetic field at a resolution of 100--1000 au (e.g.\,
\citealt{Girart2006, Attard2009, Houde2009, Hull2013, Zhang2014}), and
recently, thanks to the Atacama Large Millimeter Array (ALMA)
capabilities, it is possible to trace magnetic fields close to the
inner parts of the star forming cores (e.g.\, \citealt{Girart2018,
  Alves2018, Maury2018, Beuther2018}). However, this has been done
only in a few projects, and we need more (multiwavelength)
observations to draw a complete picture of the magnetic fields at
different evolutionary stages of the protostar.

In this paper we investigate the magnetic field of the high-mass star
forming region G9.62+0.20 by analysing ALMA observations of its dust
emission at 1 mm (band 7). We introduce G9.62+0.20 in
Sect.~\ref{sec:G9.62}.
We describe our observations and the data reduction in
Sect.~\ref{sec:observ-data-red}. In Sect.~\ref{sec:results} we present
a high-resolution polarised image of the dust emission, and we show the
morphology and the strength of the magnetic field.  Then, we compare the
results of the magnetic field direction strength obtained by our data with the
previous results obtained by masers and we discuss this comparison in
Sect.~\ref{sec:discussion}. In Sect.~\ref{sec:conclusions} we give our
conclusions and future perspectives, considering that this single case
of study can only be the starting point for a larger and more detailed
study of magnetic fields.

\section{The case of G9.62+0.20}
\label{sec:G9.62}
\object{G9.62+0.20} (hereafter G9.62) is a well-studied star forming
region, located at a distance of 5.2 kpc from the Sun
\citep{Sanna2009}, which presents several cores at different
evolutionary stages. This source contains a number of ultra-compact
\Hii regions, various masers, and hot molecular cores which drive
molecular outflows (e.g. \citealt{Persi2003, Linz2005, Liu2017},
hereafter L17 and references therein). The filament shows a
fairly well-established evolutionary sequence, making it clear that a
sequential high-mass star formation is occurring in the G9.62 complex
on a parsec-scale, from west to east \citep{Testi1998, Testi2000,
  Liu2011}.

\begin{figure*}
    \sidecaption
    \includegraphics[width=1.4\columnwidth]{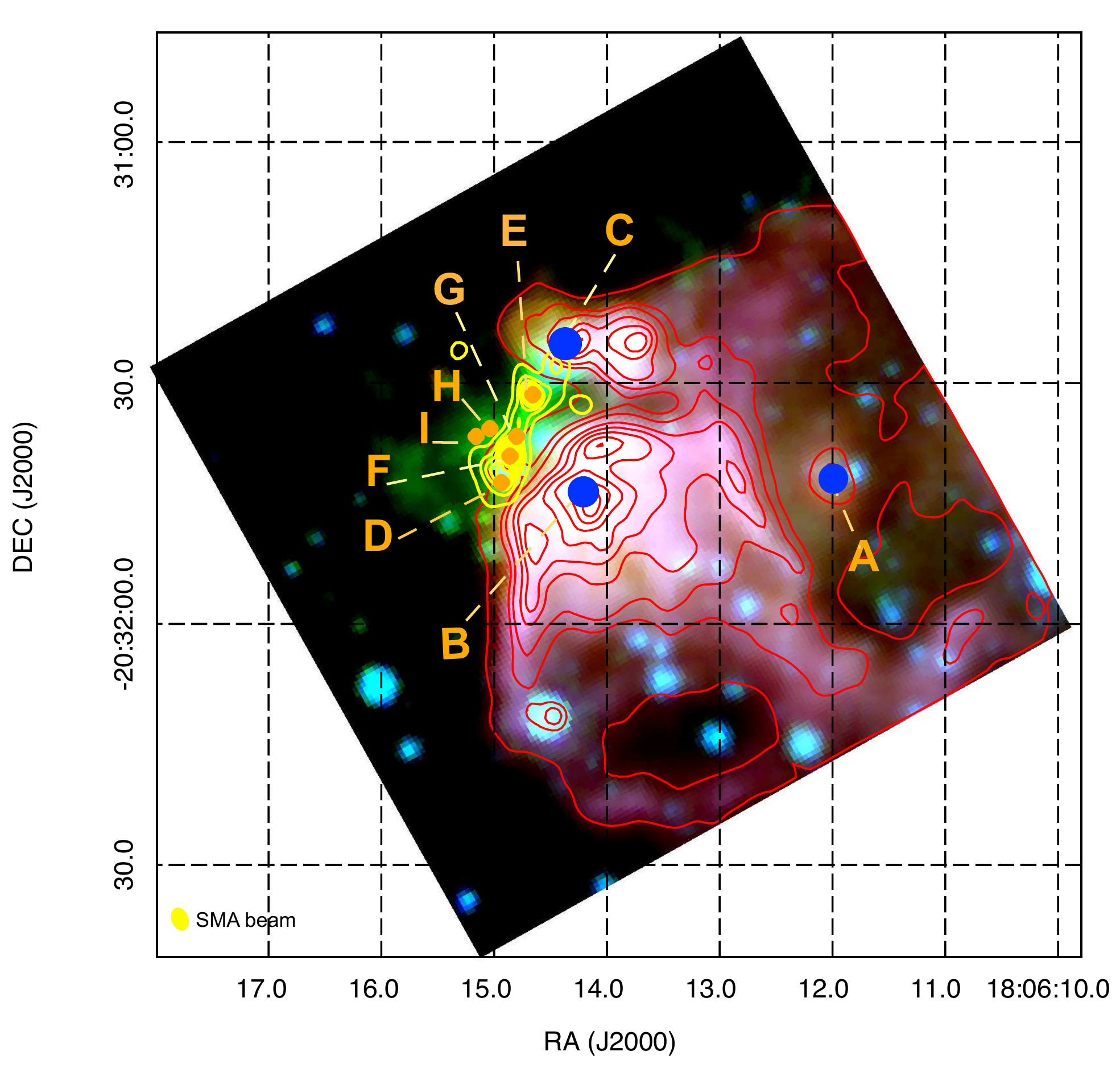}
    \caption{\textit{Spitzer}/IRAC three-colour composite image at 8,
      4.5, and 3.6 $\mu$m in red, green, and blue, respectively, of the
      star forming region G9.62. Red contours maps the
      \textit{Spitzer}/IRAC observations of PAH at 8 $\mu$m. The contours level are linearly
      from (0.1--0.9) $\times$ 2550 mJy/sr. The dark blue dots indicate
      the position of the more evolved source A, B, and C. The orange dots
 denote the positions of the massive young stellar objects denoted
      by the letters from D to I. Yellow contours track the continuum
      emission from 860 $\mu$m SMA observations of the G9.62 clump
      \citep{Liu2011}. The contours are drawn from 0.03 Jy beam$^{-1}$
      and rising in steps of 0.12 Jy beam$^{-1}$. The noise level is
      0.01 Jy beam$^{-1}$ \citep{Liu2011}.}
    \label{Fig:Spitzer}
\end{figure*}

Fig.~\ref{Fig:Spitzer} presents a \textit{Spitzer}/IRAC three-colour
composite image of the region at 8, 4.5, and 3.6 $\mu$m (in red, green,
and blue, respectively). In this figure, letters from A to I denote the
massive young stellar objects that are known centimetre and
millimetre sources. The dark blue dots indicate the positions of the evolved
young stellar objects A, B, and C, while orange dots shows the less evolved objects.
As shown in Fig~\ref{Fig:Spitzer}, the youngest sources
are located in a dense molecular clump to the east of A, B,
and C \citep{Garay1993}.

The L17 authors estimated for the G9.62 clump a mass of
$\sim2800 \pm 200$ M$_{\odot}$, a luminosity of
$\sim(1.7 \pm 0.1) \times 10^6$ L$_{\odot}$ and a mean number density
of $\sim(9.1 \pm 0.7) \times 10^4$ cm$^{-3}$. Observations by JCMT/SCUBA 
at 450 $\mu$m also traced the dust emission from the new generation of
massive protostars which are forming inside the clump and which are
surrounded by photodissociation regions (PDRs) mapped by
\textit{Spitzer}/IRAC observations of polycyclic aromatic hydrocarbon (PAH) at 8 $\mu$m (red contours in Fig~\ref{Fig:Spitzer})
\citep{Hofner2001, Testi2000, Liu2011}.

Submillimeter Array (SMA) observations at 345 GHz (860 $\mu$m)
clearly revealed multiple components \citep{Liu2011}, as shown in
Fig~\ref{Fig:Spitzer} with yellow contours. The strongest ($\sim$ 1
Jy/beam, $\sim 4.7\arcsec \times 1.3\arcsec$) is concentrated on
G9.62+0.20F. This component, with no detectable \Hii region but still
surrounded by hot dust, is in an early phase and presents signatures
of disc accretion \citep{Cesaroni1994} and it should eventually evolve
into a B0 star \citep{Testi1998}. The SMA revealed a second component
($\sim$ 0.8 Jy/beam, $\sim 1.5\arcsec \times 1.3\arcsec$) that is
concentrated on region E. The more evolved UC \Hii region G9.62+0.20E,
likely excited by a B1 star, hosts the strongest 6.7 GHz methanol
maser known \citep{Vlemmings2008, Fish2005}. The same SMA observations
also detected a smaller third core, located in the north-west part of
the filament closer to G9.62+0.20C ($\sim$ 0.2 Jy/beam, unresolved
with SMA, $\sim 1.2\arcsec$ in size) with no other related star
formation tracers. This could thus be a failed core, where the gravity
potential was unable to overcome the magnetic pressure or a remnant
core in the envelope of the UC \Hii region G9.62+0.20C
\citep{Liu2011}.

Recent ALMA observations at 230 GHz further resolved the G9.62 clump
into 12 dense cores (named MM1--MM12, L17), presenting a range of mass
spanning from 4 $M_{\odot}$ to 87 $M_{\odot}$. Five of the 12 cores are manifest
massive protostars. In L17, a possible evolutionary sequence was
proposed by the analysis of the chemistry and the outflow morphology
detected for each core, which is discussed in
Sect.~\ref{sec:discussion}.  In this paper, we refer to the entire
core $n$ as MM$n$ (e.g. MM8) and to its subcores by adding a
letter to the core name (e.g. MM8a, MM8b).

\section{Observations and data reduction}
\label{sec:observ-data-red}

G9.62 was observed with ALMA in band 7 between 13 May for $\sim$3.5
hours and 27 June 2016 for $\sim$4.5 hours (Project 2015.1.01349.S,
P.I.\ Wouter Vlemmings). The observations were performed using 39
antennas of the ALMA 12 m array in May and 42 antennas in June.

The four spectral windows with 2~GHz bandwidth each were centred at
336.5, 338.5, 348.5, and 350.5 GHz and were divided into 64 channels. The
channel width is 31.25 MHz.  At our frequency, the primary beam of
the ALMA 12 m array is 20$\arcsec$, which is large enough to encompass
all our region of interest. However, since the ALMA observatory
currently regards only the inner one-third of the primary beam as
reliable for polarisation measurements, we used two fields centred on
the main continuum sources.

The phase centres of the two fields are (at J2000) R.A. 18:06:14.558,
Dec. -20.31.30.05 and R.A. 18:06:14.889, Dec. -20.31.40.149. The two
fields are named North field and South field.  Two
different array configurations were used (C36-3, C36-4) with a total
baseline range between 17 m and 630 m in May and between 15 m and 850 m in
June\footnote{http://almascience.eso.org/observing/prior-cycle-observing-and-configuration-schedule}. The
two array configurations correspond to a maximum recoverable scale
(MRS) $\theta_{MRS}\sim2.8\arcsec$.

The datasets were reduced using the Common Astronomy Software
Application package (CASA version 4.6.0), and the calibration was
performed using J1924-2914 and J1751+0939 as bandpass calibrator,
J1733-1304 as flux calibrator, and J1832-2039 as phase
calibrator. J1924-2914 was also used as polarisation calibrator.  For
the calibration we followed the procedures provided by the ALMA
observatory. One spectral window was partially flagged before imaging
due to contamination by molecular lines (see
Sec.~\ref{sec:results}). The synthesised beam of the image is
0.34$\arcsec$ $\times$ 0.25$\arcsec$ (PA 88.53$^{\circ}$). The
1$\sigma$ rms value of the Stokes I is 0.15 mJy beam$^{-1}$, reaching
a dynamic range of 294. The continuum image suffers dynamic range
limitation because of partially resolved out flux. Comparing the flux
of the two brighter cores E and F from previous SMA observations
\citep{Liu2011}, for which the flux is extracted over a similar region in our
ALMA observations, we estimate a flux loss between 30-60\%,
respectively, on a scale larger than $\theta_{MRS}$. The absolute flux
loss of our entire region compared to single dish observations
(obtained from the ATLASGAL database \citealt{Thompson2006}) over the
same area is of the order of 70\%.

\begin{figure}
\centering
\includegraphics[width=\columnwidth, bb=263 45 628 503, clip]{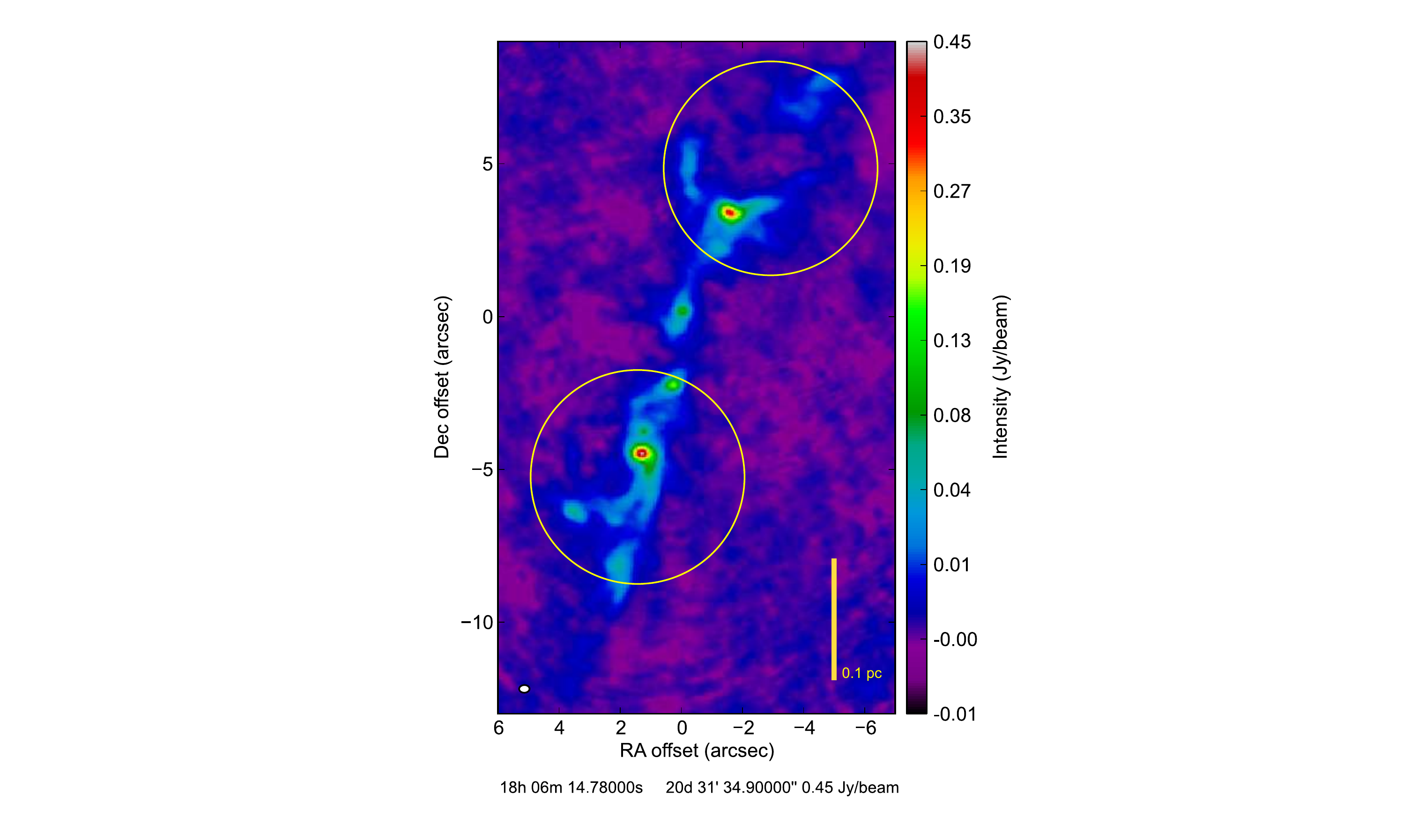}
\caption{Total intensity image of the star forming region G9.62+0.19 at
  1 mm wavelength. The yellow circles with the diameter of
  $\sim 7 \arcsec$ indicate the inner one-third of the primary beam
  for the two pointings which are overlapping. The yellow bar indicates the physical scale
  of 0.1pc, computed at the distance of the source. The colour bar
  indicates the intensity of the emission in Jy~beam$^{-1}$.The
  offsets are relative to the absolute position
  $\alpha_{2000}=18^{\rm{h}}06^{\rm{m}}14.78000^{\rm{s}}$,
  $\delta_{2000}=-20^{\circ}31'34.90\arcsec$. The white ellipse
  represents the beam.}
\label{Fig:regiontot}
\end{figure}

For the polarisation calibration, we followed the procedures provided
by the ALMA observatory. The minimum polarisation value in our data is
higher than the minimum value recommended by ALMA (0.3 \% of the total
intensity of each core, see Tab.~\ref{tab:Bpar}) and any remaining
instrumental polarisation signatures are smaller than 0.1 mJy
beam$^{-1}$.  The analysis of the linear polarisation was conducted on
each target field imaged separately, and considering only the main
substructures inside the recommended inner third of the primary beam
for the two pointings, as shown in Fig.~\ref{Fig:regiontot}. The
1$\sigma$ rms value of the linear polarised image is on average
$\sigma_{P}$=0.08 mJy beam$^{-1}$, obtained using
$\sigma_{P}=
\sqrt{[(Q\times\sigma_Q)^2+(U\times\sigma_U)^2]/(Q^2+U^2)}$ and we
conducted the polarisation analysis above a signal-to-noise ratio of
4. This selection was applied to perform the analysis of the polarised
emission and the magnetic field only in the proximity of the cores. We
combined the U and Q data cubes to produce cubes of polarised intensity
(I$_P$=$\sqrt{Q^2+U^2-\sigma_{P}^2}$) and polarisation angle
($\psi=1/2 \times \mathrm{atan} (U/Q)$). The error on polarisation
angles includes the formal error due to the thermal noise
\citep{Wardle&Kronberg1974} and this error is given by
$\sigma_{tn} = 0.5(\sigma_{P}/$I$_P)\times(180^{\circ}/\pi)$. For the
4$\sigma$ cut this value corresponds to 7.2 degrees and for the
strongest peaks this decreases to 0.7 degrees.

\section{Results}
\label{sec:results}
\subsection{Detection of the cores}
\label{sec:cores}

\begin{figure*}
 \centering
  \includegraphics[width=0.4\textwidth, bb=263 45 628 503, clip]{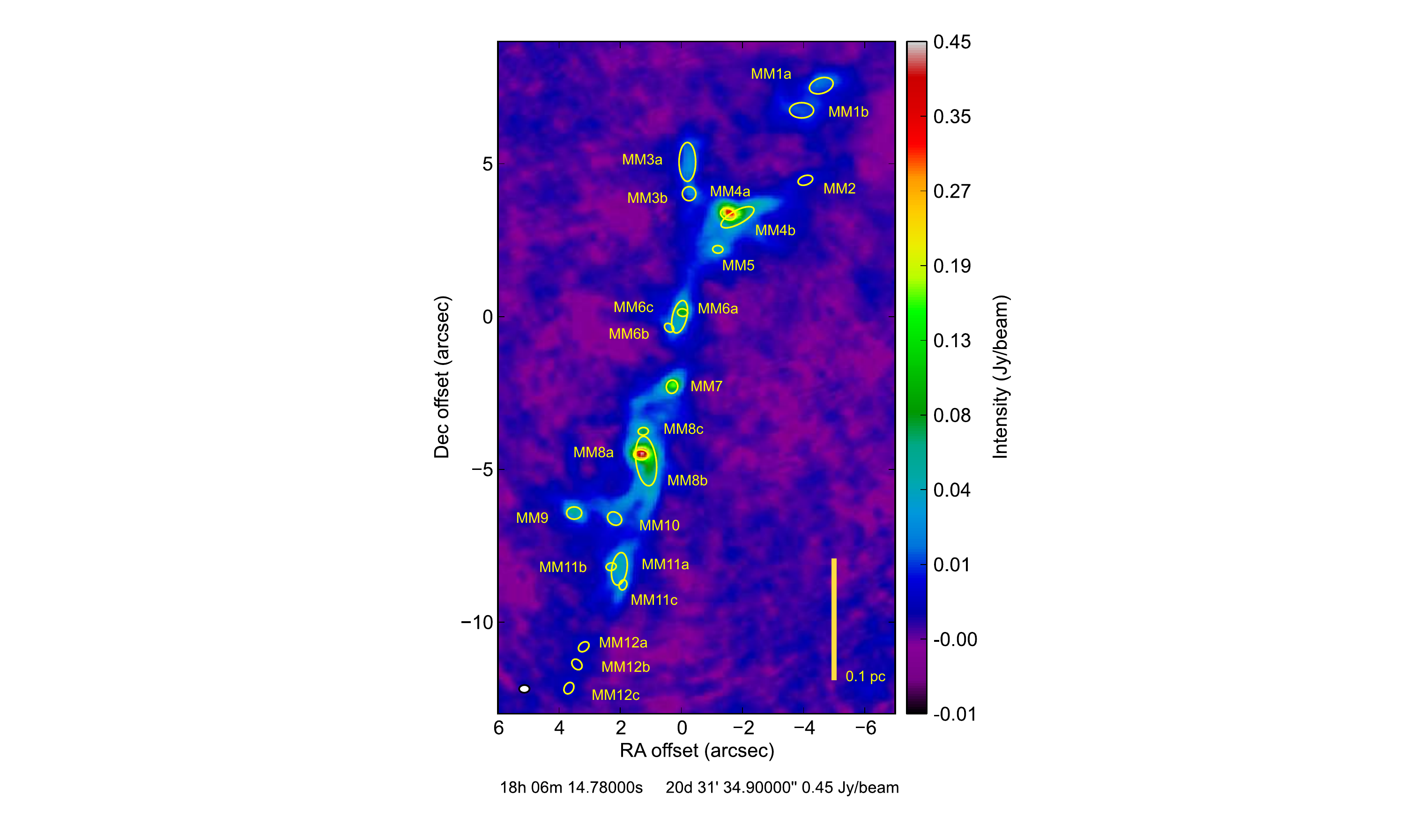} 
  \includegraphics[width=0.4\textwidth, bb=263 45 628 503, clip]{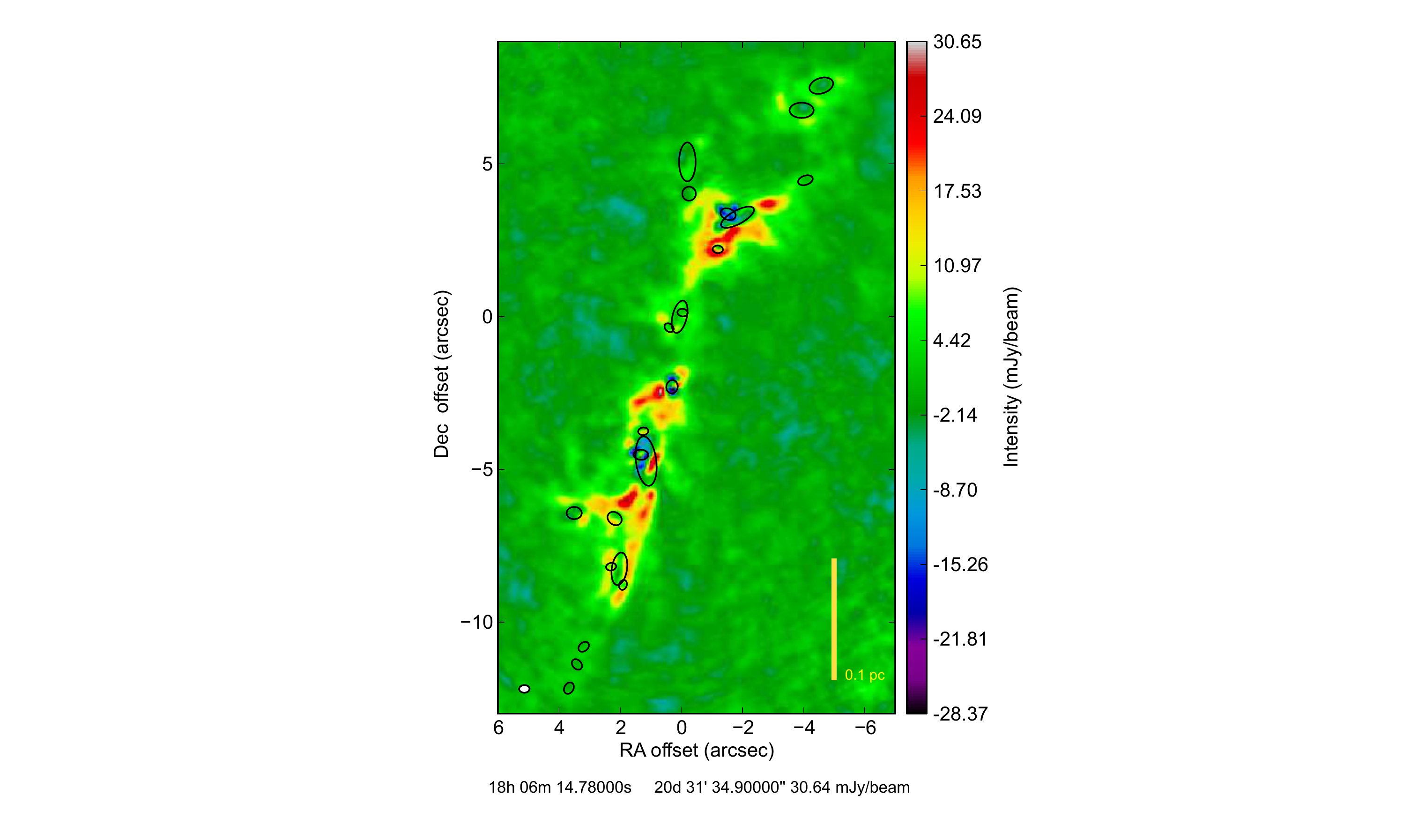}
  \caption{Left panel: total intensity image of the star forming region
    G9.62+0.19, with the identified cores. The offsets are relative to
    the absolute position
    $\alpha_{2000}=18^{\rm{h}}06^{\rm{m}}14.78000^{\rm{s}}$,
    $\delta_{2000}=-20^{\circ}31'34.90\arcsec$. The white ellipse
    represents the beam and the yellow ellipses represent the dense
    cores identified by the Gaussian fit and illustrated in
    Tab.~\ref{tab:continuum_sources}. The colour scale goes from -0.01
    to 0.45 Jy beam $^{-1}$.  Right panel: residual image in which the black ellipses
    represent the dense cores as before. The colour scale goes from
    -28.37 to 30.65 mJy beam $^{-1}$. In both panels, the yellow bar indicates
    the physical scale of 0.1 pc, at the distance of the source.}
 \label{Fig:mosaico1}
\end{figure*}

In Fig.~\ref{Fig:regiontot} we present the Stokes I image of the G9.62
clump as seen in our ALMA band 7 continuum observations. The two
yellow circles outline the inner third of the two pointings that we
observed in this project. The filament presents a highly fragmented
structure, extending from the north-west to the south-east direction. We
confirm the presence of the 12 dense cores, including the main three
continuum peaks MM3, MM4, and MM8, already seen in the previous ALMA
band 6 observations by L17. In addition, since our ALMA band 7
observations have about three times higher angular resolution than L17, we
identify several other substructures that have signal-to-noise ratios above 40 $\sigma$. In
Tab.~\ref{tab:continuum_sources} we list 23 cores and their
characteristics, and we identify these in Fig.~\ref{Fig:mosaico1}. The positions, peak flux
densities, integrated fluxes, and sizes of the cores are obtained from
two-dimensional Gaussian fits performed with CASA task IMFIT. We observe the
presence of large amount of extended emission that we model using
single Gaussian and point-like sources.  This analysis is based on the
combined dataset of both observed fields.

\begin{sidewaystable*}[p!] 
\caption []{Parameters of the continuum sources of the 1 mm ALMA observations of G9.62+0.19.} 
\begin{center}
\setlength{\tabcolsep}{4.0pt}
\begin{tabular}{lccccccccccccc}
\hline\hline
 core& & RA \tablefootmark{a} & Dec \tablefootmark{a}    & $I_{Peak}$   \tablefootmark{b} & Integrated flux  \tablefootmark{c} & a   \tablefootmark{d}      & b  \tablefootmark{d}     & P.A.  \tablefootmark{d}  &  $\alpha$  \tablefootmark{e}& $\beta$ \tablefootmark{f}& N(H$_2$)\tablefootmark{f}& Mass\tablefootmark{f} & $\lambda$\tablefootmark{g} \\
     & & Offset (“)           & Offset  (“)              & (mJy/beam)                     & (mJy)                              & (milliarcsec)              & (milliarcsec)            & $\circ$                &                              &                          & cm$^{-2}$               & M$_\odot $             &   \\
\hline                                                                                                                                                                                                              
MM1a & & -4.58  $\pm$ 0.04    & 7.56       $\pm$    0.02 & 24.5      $\pm$          2.7   & 114.0    $\pm$     13.0            & 726       $\pm$        4   & 426       $\pm$       7  & 110.6  $\pm$    9.6      & 3.4 $\pm$ 0.5              &  1.6    &2.9$\times$10$^{24}$  & 10& --- \\
MM1b & & -3.93  $\pm$ 0.05    & 6.75       $\pm$    0.02 & 21.4      $\pm$          2.7   & 98.0     $\pm$     12.0            & 712       $\pm$        2   & 432       $\pm$       3  & 90     $\pm$    12       & ---          &---&---&---&---\\
MM2  & & -4.06  $\pm$ 0.00    & 4.46       $\pm$    0.00 & 11.0      $\pm$          0.0   & 19.2     $\pm$     0.0             & 374       $\pm$        13  & 138       $\pm$       42 & 117    $\pm$    16       &  ---         &--- &---&---&---\\
MM3a & & -0.20  $\pm$ 0.00    & 5.06       $\pm$    0.00 & 27.6      $\pm$          2.0   & 221.0    $\pm$     18.0            & 1251      $\pm$        105 & 419       $\pm$       47 & 0.06  $\pm$    0.02    &  3.3 $\pm$ 0.3             &  1.5    &3.5$\times$10$^{24}$& 21& 8\\
MM3b & & -0.25  $\pm$ 0.00    & 4.02       $\pm$    0.00 & 23.0      $\pm$          2.7   & 54.4     $\pm$     8.8             & 384       $\pm$        69  & 281       $\pm$       93 & 1.16   $\pm$    0.68     & ---          &---&--- &---&---\\
MM4a &E & -1.53  $\pm$ 0.00    & 3.35       $\pm$    0.00 & 321.6     $\pm$          6.8   & 699.0    $\pm$     20.0            & 402       $\pm$        18  & 238       $\pm$       15 & 58.4   $\pm$    4.5      & 3.8 $\pm$ 0.2             &  1.8   &9.5$\times$10$^{24}$&43& 6\\
MM4b &E & -1.84  $\pm$ 0.03    & 3.25       $\pm$    0.02 & 82.6      $\pm$          5.3   & 508.0    $\pm$     37.0            & 1162      $\pm$        94  & 345       $\pm$       34 & 119.4  $\pm$    2.1      & ---          &--- &---& ---&---\\
MM5  & & -1.19  $\pm$ 0.02    & 2.20       $\pm$    0.01 & 43.1      $\pm$          6.2   & 42.4     $\pm$     6.1             & ---                        & ---                       & ---                       & ---         &--- &--- &---&--- \\
MM6a & & -0.04  $\pm$ 0.00    & 0.13       $\pm$    0.00 & 65.6      $\pm$          1.9   & 64.6     $\pm$     1.8             & ---                        & ---                       & ---                      & ---          &---&---&---&---\\
MM6b & & 0.40   $\pm$ 0.02    & -0.37      $\pm$    0.02 & 14.8      $\pm$          1.9   & 14.6     $\pm$     1.8             & ---                        & ---                       & ---                      & ---          &---&---&---&---\\  
MM6c & & 0.06   $\pm$ 0.01    & -0.01      $\pm$    0.02 & 42.6      $\pm$          1.6   & 251.0    $\pm$     11.0            & 1057      $\pm$        50  & 324       $\pm$       24  & 167.3  $\pm$    1.5        & 3.9 $\pm$ 0.3            &  2.1   &5.8$\times$10$^{24}$&22 & ---\\
MM7  &G & 0.30   $\pm$ 0.01    & -2.30      $\pm$    0.01 & 168.9     $\pm$          11.0  & 321.0    $\pm$     30.0            & 359       $\pm$        45  & 152       $\pm$       82 & 175    $\pm$    15       & 3.8 $\pm$ 0.5             &  1.8   &3.5$\times$10$^{25}$&15& ---\\
MM8a &F & 1.32   $\pm$ 0.00    & -4.52      $\pm$    0.00 & 366.7     $\pm$          7.5   & 675.0    $\pm$     20.0            & 342       $\pm$        17  & 210       $\pm$       11 & 85.2   $\pm$    4.1      & 3.8 $\pm$ 0.3             &  1.8   &4.7$\times$10$^{25}$&38& 21\\
MM8b &F & 1.15   $\pm$ 0.02    & -4.73      $\pm$    0.04 & 80.7      $\pm$          4.8   & 1011.0   $\pm$     65.0            & 1612      $\pm$        107 & 568       $\pm$       44 & 7.2    $\pm$    2.2      & ---           &---&---& ---&---\\
MM8c &F & 1.25   $\pm$ 0.03    & -3.75      $\pm$    0.01 & 44.6      $\pm$          6.9   & 43.9     $\pm$     6.8             & ---                        & ---                      & ---                      & ---           &---&---& ---&---\\
MM9  & & 3.51   $\pm$ 0.03    & -6.43      $\pm$    0.02 & 63.4      $\pm$          7.6   & 147.0    $\pm$     17.0            & 364       $\pm$        0   & 310       $\pm$       0   & 87.8   $\pm$    0.1         & 3.1 $\pm$ 0.7            &  1.3  &1.9$\times$10$^{25}$& 23& ---\\
MM10 & & 2.18   $\pm$ 0.04    & -6.61      $\pm$    0.03 & 23.0      $\pm$          4.1   & 53.4     $\pm$     9.6             & 402       $\pm$        34  & 259       $\pm$       47  & 36     $\pm$    36        &  ---         &---&--- &---&---\\
MM11a &D& 2.03   $\pm$ 0.01    & -8.26      $\pm$    0.03 & 42.8      $\pm$          2.1   & 277.0    $\pm$     15.0            & 1046      $\pm$        62  & 363       $\pm$       32 & 172.1  $\pm$    2.1      & 2.7 $\pm$ 0.4             &  0.9   &8.7$\times$10$^{24} $&37& ---\\
MM11b& D& 2.30   $\pm$ 0.02    & -8.19      $\pm$    0.01 & 20.3      $\pm$          2.4   & 20.0     $\pm$     2.4             & ---                        & ---                      & ---                      & ---        & ---  &--- &---&---\\
MM11c &D& 1.91   $\pm$ 0.01    & -8.78      $\pm$    0.02 & 17.3      $\pm$          2.4   & 17.1     $\pm$     2.4             & ---                        & ---                      & ---                      & ---       & ---   &--- &---&---\\
MM12a && 3.20   $\pm$ 0.03    & -10.81     $\pm$    0.03 & 6.8       $\pm$          1.4   & 8.7      $\pm$     1.8             & ---                        & ---                       & ---                    &   ---    &--- &---&---&---\\
MM12b& & 3.42   $\pm$ 0.03    & -11.38     $\pm$    0.04 & 6.2       $\pm$          1.4   & 8.3      $\pm$     1.9             & ---                        & ---                       & ---                    &   ---    &---&---&---&---\\
MM12c && 3.68   $\pm$ 0.03    & -12.16     $\pm$    0.03 & 7.3       $\pm$          1.4   & 10.3     $\pm$     2.0             & ---                        & ---                       & ---                    &   ---    &---&---&--- &---\\
\hline
\end{tabular}
\end{center}
\tablefoot{ \tablefoottext{a}{The offsets are relative to the absolute
    position $\alpha_{2000}=18^{\rm{h}}06^{\rm{m}}14.78^{\rm{s}}$
    $\delta_{2000}=-20^{\circ}31'34.9\arcsec$.}
  \tablefoottext{b}{Peak flux density.}  \tablefoottext{c}{Total flux
    density from the 2D Gaussian fit. To model MM6, MM8, and MM11 we
    make use of Gaussian components and point-like source
    components. CASA task \textit{imfit} was used to perform
      the Gaussian fit.}  \tablefoottext{d}{$a$, $b,$ and $P.A.$ are
    the deconvolved FWHM major and minor axes and position angle,    obtained from 2D Gaussian fits. Point sources are those for which
    $a$, $b,$ and $P.A.$ are missing. Uncertainties of 0.00 and 0.0 are
    $<0.01$ and $<0.1,$ respectively.}  \tablefoottext{e}{Spectral
    index $\alpha$ obtained between our ALMA band 7 observations and
    ALMA band 6 observations by Liu et al.\ 2017.}
  \tablefoottext{f}{The spectral index of the dust opacity $\beta$,
    the column density N(H$_2$), and the mass in solar units are
    estimated considering a temperature of the dust T$_d\sim$35 K for
    the starless core MM1a, MM3a, MM9, and for the young sources MM6c
    and MM11a, while a T$_d\sim$100 K was used for the hot cores MM4a,
    MM7 and MM8a.}  \tablefoottext{g}{To compute the mass to
      flux ratio we used the magnetic field values obtained by SF analysis.}  }
\label{tab:continuum_sources}          
\end{sidewaystable*}

\subsection{Spectral index of the dust emission}
\label{sec:spectral-index}

In Tab.~\ref{tab:continuum_sources} we also give the values of the
spectral index $\alpha$, computed considering the integrated
intensities of the cores from ALMA band 6 observations (L17) and
similar visibility coverage to our observations. For a relevant
measurement of the spectral index, each of the spectral windows in the
two datasets were imaged (line free channels) using only the
overlapping $uv$ distances (20--360~k$\lambda$) and restoring the images
with a circular beam of 0.9$\arcsec$. While this limits the comparison
to the peaks given in L17, it can hint at the evolutionary status of
the cores which can then be compared to the polarisation images.

The dust emission at submillimetre wavelengths can be represented by a
grey body, with intensity varying smoothly as a function of
frequency. Having measured the intensity of dust emission at two submillimetre
frequencies, it is possible to describe the spectral energy
distribution (SED) of each core using the spectral index of the dust
emission $\alpha$ defined as

\begin{equation}
\alpha = \frac{\ln (S_a/S_b)}{\ln(\nu_b/\nu_a)}  \quad,\end{equation}

where $S_a$ and $S_b$ are flux densities at frequencies $\nu_a$ and
$\nu_b$, respectively \citep{Williams2004}.

Within the errors all the cores present roughly the same
  spectral index. Since $\alpha$ depends on a combination of several
  factors, the uncertainties are influenced by the errors on the
  opacity, beam-averaged dust temperature, and spectral
  index of the dust opacity ($\beta$, defined as
  $\tau \propto \nu^\beta$). Also the absolute flux calibration uncertainty
  is a significant source of error, although this is a
  systematic error in the spectral index for all regions that can be
  up to ~0.2 (based on~10\% flux calibration uncertainties).
Moreover, changes in the composition of the dust
grains and temperature variations through the dust envelope can
influence $\alpha$ . The models developed by \citet{Ossenkopf1994}
predicted that $\beta$  changes if the dust grains do not have ice
mantles; this should be the case of some of our cores, since they are
in the hot core phase reaching inner dust temperatures higher than 100
K.  This fact is also confirmed by the detection of CH$_3$OH and
CH$_3$CN, well-known molecular tracers of hot cores \citep{Qiu2014},
pointing to high temperature and high density regions where the
environment is warm enough to melt ice mantles and permit grain growth
\citep{Williams2004}.

We can use an estimate of the temperature of dust grains within the
cores to calculate the spectral index of the dust opacity
\begin{equation}
 \beta=(\alpha+\Delta\alpha)-2  \quad,
\label{eq:beta}
\end{equation}
where $\Delta\alpha$ is defined as the
Rayleigh-Jeans correction factor \citep{Williams2004}.

The dust temperatures are closer to the equivalent temperature at a
given frequency $T_\nu=h\nu/k$ than a Rayleigh-Jeans approximation
would permit, where $h$ is the Planck constant and $k$ is the Boltzmann
constant in cgs unit.  Thus, for frequencies $\nu_a=230$ GHz and
$\nu_b=340$ GHz in ALMA band 6 and band 7, we obtain $T_{\nu_a}= 11$ K
and $T_{\nu_b}=16$ K, respectively, from which we can estimate the
correction $\Delta\alpha$ as
\begin{equation}
\Delta\alpha = \frac{\ln \left(\frac {e^{T_{\nu b}/T_\mathrm{d}} -1}{e^{T_{\nu a}/T_\mathrm{d}} -1}\right)}{\ln(T_{\nu_b}/T_{\nu_a})} -1 \quad.
\label{eq:delta-alfa}
\end{equation}

Assuming a dust temperature of $T_\mathrm{d1}\sim100$ K and
$T_\mathrm{d2}\sim$ 35 K in the hot cores and in the starless cores,
respectively (values proposed by L17), we obtain from Eq.~(\ref{eq:delta-alfa})
$\Delta\alpha_1$= 0.06 and $\Delta\alpha_2$=0.20 in the two
environments.  We then estimate the spectral index of the dust opacity
$\beta$ using Eq.~(\ref{eq:beta}), and report all the values
in Tab.~\ref{tab:continuum_sources}. Since the error on $\Delta\alpha$
depends on the uncertainty on $T_\mathrm{d}$, which were not estimated by
L17, the error on $\beta$ has the same order of uncertainties of
the spectral index $\alpha$ and could be underestimated.

\subsection{Column densities and masses of the cores}
\label{sec:spectral-index}

Given the above values for the dust temperature T$_d$, and following the
approach described in L17, we can estimate the H$_2$
column densities (N$_{H_2}$) and gas masses for the regions.
The integrated flux $S_\nu$ from thermal dust is
\begin{equation}
\label{eq:flusso}
S_\nu = B_\nu(T_d) \left[ 1-e^{-\tau_d} \right] \Omega_s
,\end{equation}
where the observed frequency $\nu$ is expressed in Hz, $\Omega_s$ is the solid angle in steradiants, and $\tau_d$ is the optical depth. From the Planck function
\begin{equation}
\label{eq:plankiana}
B_\nu(T_d) = \frac{2h\nu^3}{c^2} \frac{1}{e^{\frac{h\nu}{kT_d}}-1}
,\end{equation}

where c is the velocity of light in cm s$^{-1}$. Considering a dust temperature of $T_\mathrm{d1}\sim100$ K and
$T_\mathrm{d2}\sim$ 35 K in the hot cores and in the starless cores, we can derive from Eq.~(\ref{eq:plankiana}) the optical depth $\tau_d$ and we can compute the mass of the dust $\mathrm{M_d}$, which is defined as

\begin{equation}
\label{eq:mass-dust}
\mathrm{M_d}= \frac{\Omega_s d^2 \tau_d}{ \kappa_\nu}   
,\end{equation}

where $\kappa_\nu$ is the dust opacity per unit dust mass and  $d$ is the distance in cm. We derived the
dust opacity index $\kappa_\nu$ from \citet{Ossenkopf1994} for grains
with ice mantles using
\begin{equation}
\label{eq:kappa}
\kappa_\nu = \kappa_0 \left( \frac{\nu}{\nu_0} \right)^\beta  \quad.
\end{equation}

We also take into account two different density regimes for dense hot
cores and for starless cores, as reported in L17.  In the first case
we adopted an average number density $n=$10$^6$ cm$^{-3}$ for the denser cores MM4, MM7,
and MM8 and $\kappa_{300\mathrm{GHz}}\sim$ 1.37 cm$^2$ g$^{-1}$
provided by \citet{Ossenkopf1994}. In the second case, we assumed
$n=$10$^5$ cm$^{-3}$ for starless cores MM1a, MM3a, MM9, and for the
young sources MM6c and MM11a and we derived $\kappa_{\nu}$ from
$\kappa_{230\mathrm{GHz}}\sim$ 0.9 cm$^2$ g$^{-1}$ given by
\citet{Ossenkopf1994}.
We consider a gas-to-dust mass ratio of 100 and thus we obtained
the column densities from

\begin{equation}
\label{eq:NH2}
N(H_2)=\frac{ M_g}{\tau_d}{ \mu m_H \Omega_s d^2}
,\end{equation}  
where $\mathrm{M_g}$ is the mass of the gas.

The derived values are presented in Tab.~\ref{tab:continuum_sources}:
the core masses range between roughly 10 M$_\odot$ and 40 M$_\odot$ and the
column densities between 2 $\times$ 10$^{24}$ and 6 $\times$ 10$^{25}$
cm$^2$, which are a factor of 10-100 larger than the mean values
reported in L17. The optical depth ranges typically between
  0.2 and 0.9 with uncertainties of $\sim$20\%, mainly determined by the
  errors on the dust model and on the radii from the Gaussian fits, spectral index, and temperature. For sources MM7 and MM8
  the errors on the optical depth are much larger and we can only
  constrain $\tau_d < 3.5$. The uncertainties on the optical depth may
  cause an overestimation of the mass if the emission becomes
  optically thick. Therefore considering all these uncertainties, the
  errors of the masses and column densities are difficult to quantify
  and they could be more than a factor of 5 and less than a factor of 10.

\subsection{Detection of linear polarisation and magnetic field morphology}
\label{sec:polarisation}

\begin{table*}
  \caption[]{Polarised intensities and magnetic field parameters.}
  \label{tab:Bpar}
  \centering
  \begin{tabular}{lccccccccccc}
    \hline\hline
     Core & I$_{peak}$          & I$_{pol}$\tablefootmark{a}                 & $\sigma_{\psi}$\tablefootmark{b}       & $\sigma_{\nu}$\tablefootmark{c}   & $B_{\perp}^{DCF}$       &b                        & $B_{\perp}^{SF}$ & $B_t/B_0$\\
          & (mJy beam$^{-1}$)   & (mJy beam$^{-1}$)       & ($^{\circ}$)                          &(km s$^{-1}$)                      & (mG)                 & ($^{\circ}$)                 & (mG)          &           \\

    \hline
    MM3       & 27.6         & 2.79$\pm$0.07              & 41.8  $\pm$ 0.1          & 2.00 $\pm$ 0.01                            &        0.33  $\pm0.02$            &11.9  $\pm$ 0.2       &   3.30  $\pm0.07$  &     0.2   \\
    MM4       & 321.6        & 1.80$\pm$0.06              & 28.0  $\pm$ 0.4          & 2.00  $\pm$ 0.01                           &        1.57  $\pm0.02$            &9.4  $\pm$ 0.3        &   13.18 $\pm0.42$ &     0.1   \\
    MM7       & 168.9        & 0.66$\pm$0.08              & ---                      & ---                                        & ---                                &---                  & ---               &   ---    \\
    MM8a      & 366.7        & 1.74$\pm$0.08              & 30.5  $\pm$ 3.8          & 2.35  $\pm$ 0.01                           &        1.64  $\pm0.02$            &8.6 $\pm$ 0.4         & 17.10 $\pm0.75$ &     0.1    \\
    MM8c      & 44.6         & 1.25$\pm$0.08              & ---                      & ---                                        & ---                                &---                  & ---      &    ---     \\
    MM9       & 63.4         & 0.91$\pm$0.07              & ---                      & ---                                        & ---                                &---                  & ---      &   ---     \\
    MM11a     & 42.8         & 0.54$\pm$0.10              & ---                      & ---                                        & ---                                &---                  & ---      &   ---     \\                                                                                  
    \hline   
  \end{tabular}
  \tablefoot{
    \tablefoottext{a}{I$_{pol}$ are taken at the peak of the polarised intensity. In case of MM3 and MM4 the peak is located in MM3a and MM4a, respectively.}
    \tablefoottext{b}{All the values of $\sigma_{\psi}$ are obtained using the classical variance, calculated around the circular mean \citep{jammalamadaka2001,pewsey2013}, over regions covering MM3, MM4, and MM8. }
    \tablefoottext{c}{All the values of $\sigma_{\nu}$ are obtained by Liu et al. 2017, Table 3, using methanol emissions.}
    }
\end{table*}

\begin{figure*}
 \centering
  \includegraphics[width=\columnwidth, bb=128 60 700 484, clip]{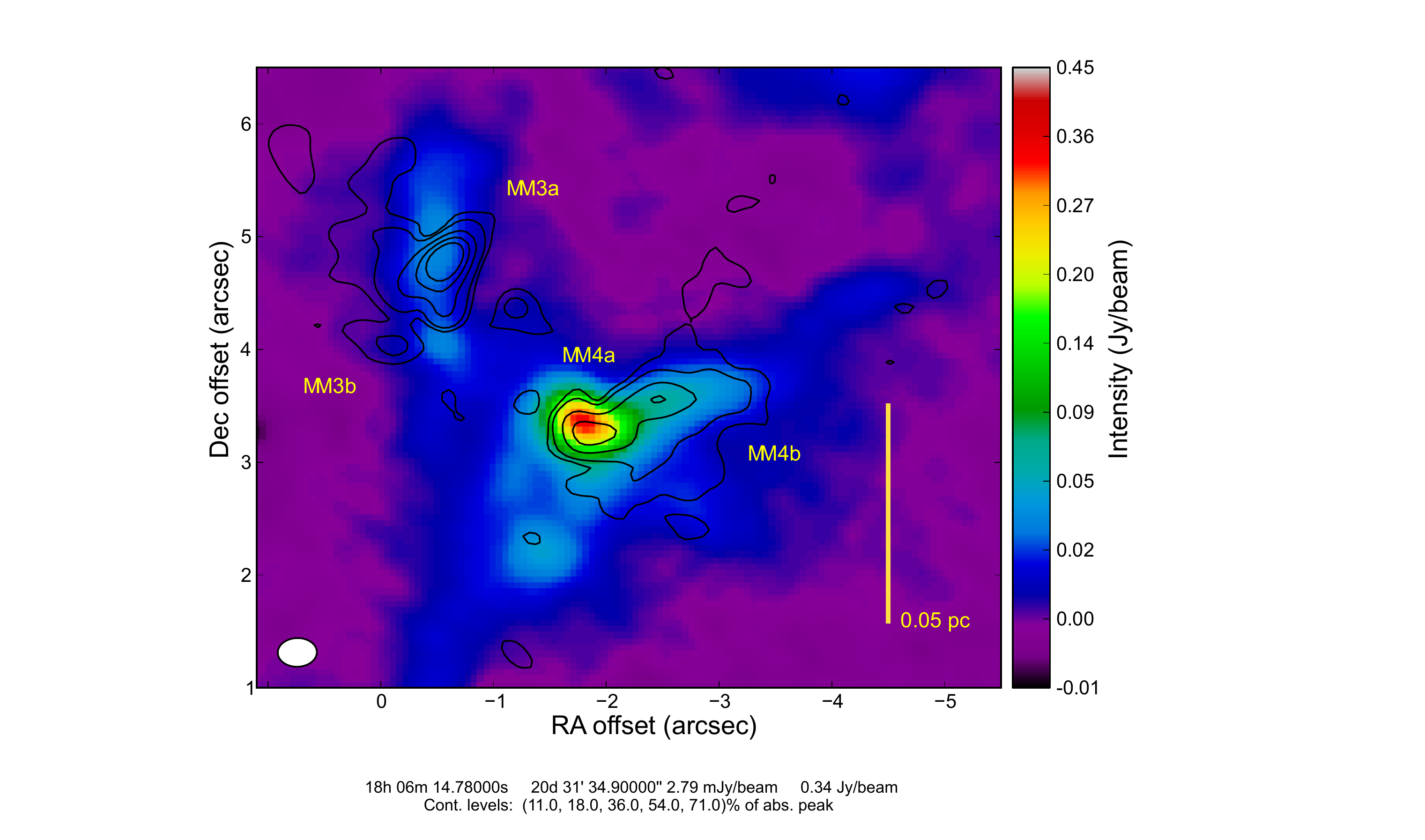}
  \includegraphics[width=\columnwidth, bb=128 60 700 484, clip]{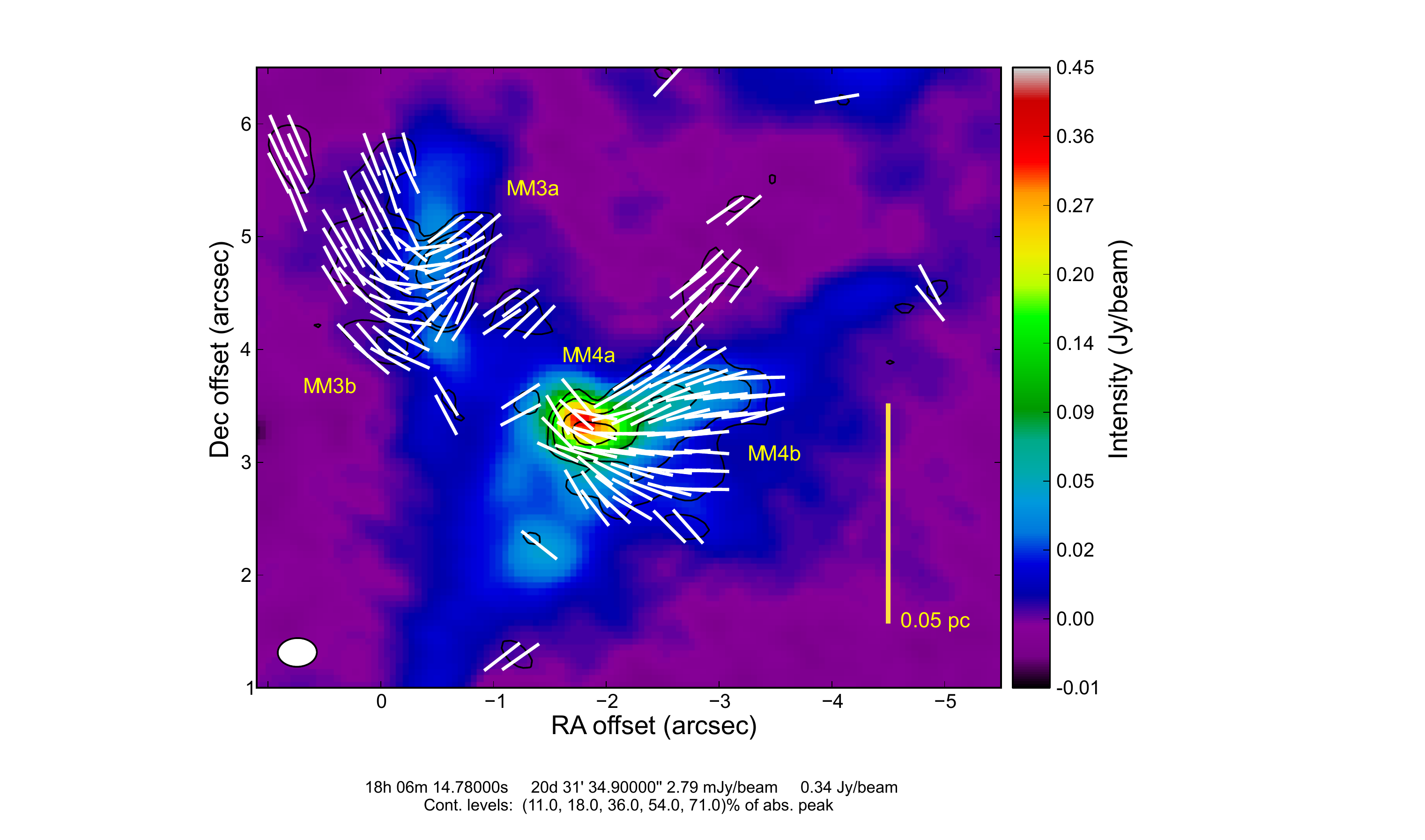} 
  \caption{Zoom on the cores MM3 and MM4 with the relative subcores.      
      Left panel: the contours represent linearly polarised emission and the
      levels are (0.3, 0.5, 1.0, 1.5, 2.0) mJy beam$^{-1}$.
      Right panel: the segments indicate linear polarisation vectors
      already rotated of 90$^{\circ}$ to show the orientation of the
      magnetic field. Only in this figure, the sampling of the vectors are every 165 by 165 
      mas. In both, the offsets are relative to the same
      position as in Fig.~\ref{Fig:mosaico1} and the colour scale
      indicates the total intensity of the background image in Jy beam $^{-1}$, going from
      -0.01 to 0.45 Jy beam $^{-1}$; the yellow bar indicates the physical scale of
      0.05 pc, at the distance of the source.}
 \label{Fig:mosaicozoomN}
\end{figure*}

\begin{figure*}
 \centering
 \includegraphics[width=.9\columnwidth, bb=244 56 559 491, clip]{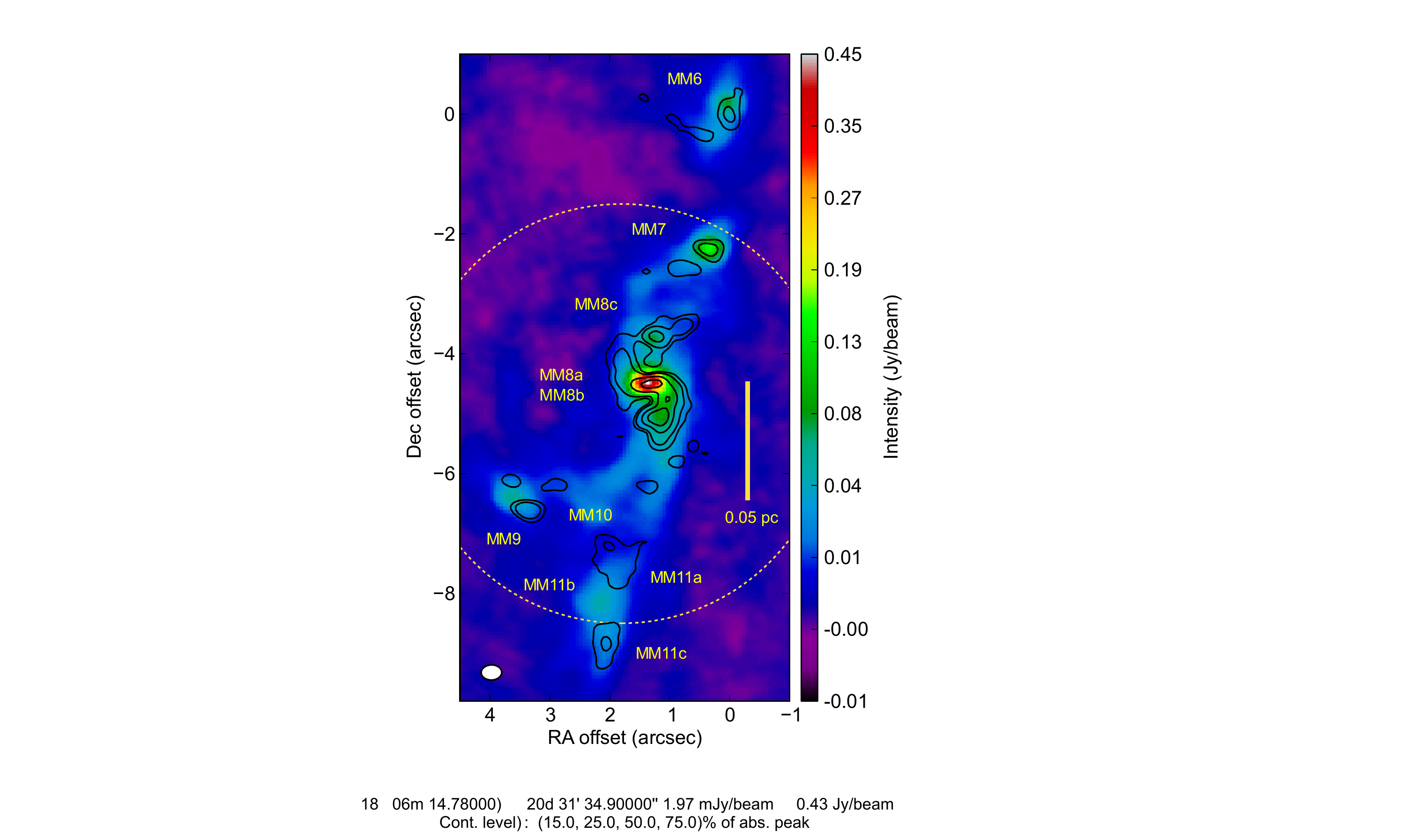}
 $\quad$
 \includegraphics[width=.9\columnwidth, bb=244 56 559 491, clip]{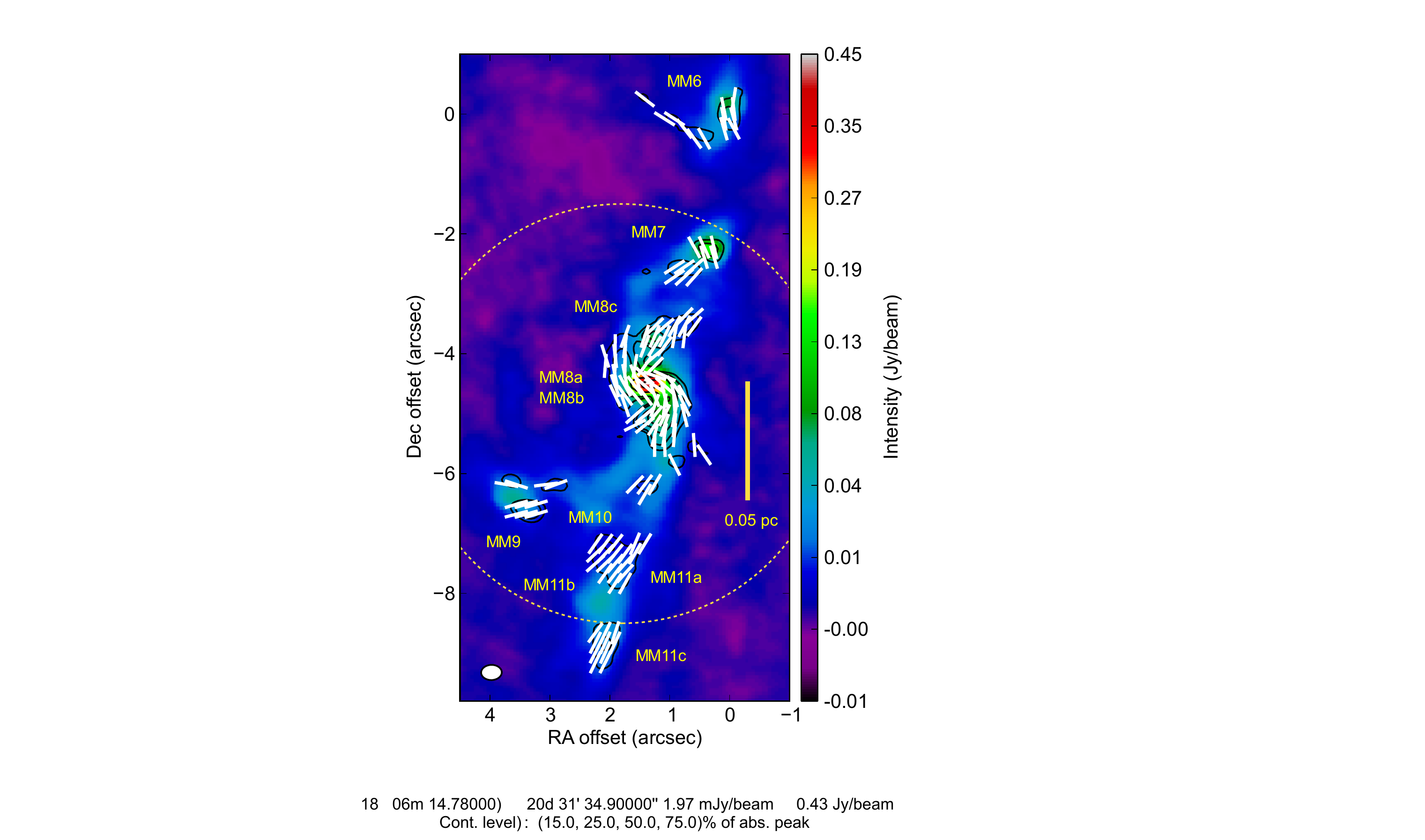}
 \caption{Zoom on the core MM8 and relative subcores. Left panel: the
   contours represent linearly polarised emission and the levels are
   (0.3, 0.5, 1.0, 1.5, 2.0) mJy beam$^{-1}$. Right panel: the segments indicate
   linear polarisation vectors already rotated of 90$^{\circ}$ to show
   the orientation of the magnetic field. Only in this figure, the sampling of the vectors
   are every 165 by 165 mas. In both, the offsets are relative to the same
   position as in Fig.~\ref{Fig:mosaico1} and the colour scale is
   relative to the background image that indicates the total intensity
   in Jy beam $^{-1}$, going from -0.01 to 0.45 Jy beam $^{-1}$; the
   yellow bar indicates the physical scale of 0.05 pc, at the distance
   of the source. The dotted yellow circle indicates the inner one-third of the primary beam (as already described in
  Fig~\ref{Fig:regiontot}).}
 \label{Fig:mosaicozoomS}
\end{figure*}

We detect linear polarised emission in six regions: two in the
northern field (from MM3 and MM4) and four in the southern field (from
MM7, MM8, MM9 and MM11a). The total intensity (I$_{peak}$) and linear
polarised intensity (I$_{pol}$) are reported in
Tab.~\ref{tab:Bpar}. The linear polarised emission coming from the
north-west source MM3 is particularly strong, and the polarised signal
at its peak is 10\% of the total intensity. This value must be
considered as an upper limit because of the flux loss, which is higher
in the stokes I than in the polarised signal. A zoom-in of the four
cores is given in Fig.~\ref{Fig:mosaicozoomN} and
Fig.~\ref{Fig:mosaicozoomS}, where the black contours represent the
linear polarised emission and the white segments denote the magnetic
field, where we assumed that the magnetic field is perpendicular to
the linear polarisation vectors.The sampling of the vectors, only in these figures,
   are every 165 by 165 mas. At our densities and at the given
scales the polarised signal is dominated by magnetic field effects and
not by scattering \citep{Kataoka2016,Kataoka2017}. The polarisation
vectors in each of the six cores exhibit an ordered structure
indicating that the magnetic field may be important in these sources,
because, in case of turbulence dominated regions, the polarisation
vectors distribution would have been less ordered.  While in all the
cores in MM3 and MM4, and in MM7, MM8c, MM9, and MM11a the polarisation
vectors show smooth angle shifts, in the more extended region
including the subcores MM8a and MM8b the polarisation angle
distribution resembles a spiral structure. Nevertheless, the magnetic
field morphology of the whole region seems to follow the direction of
the stream of dust along the filament. In Fig.~\ref{Fig:mosaicozoomS}
the dotted yellow circle indicates the inner one-third of the primary
beam (as already described in Fig~\ref{Fig:regiontot}). Following ALMA
recommendations we have excluded from our analysis the polarised
emission coming from MM6 and MM11c, since they are located beyond the
one-third of the primary beam.

\subsection{Magnetic field strength}
\label{sec:magneticfield}
\subsubsection{Davis-Chandrasekhar-Fermi method}
\label{sec:structure_function}
The magnetic field strength in a gas can be inferred
by applying the Davis-Chandrasekhar-Fermi (DCF) method
\citep{Davis1951,Chandrasekhar-Fermi1953}, following the procedure
described in \citet{Planck_coll2016}. Given the volume gas density $\rho$,
the angular dispersion of the local magnetic field orientation
$\sigma_{\psi}$ and the one-dimensional velocity dispersion of the gas
$\sigma_{\nu}$, the strength of the magnetic field component in the
plane of the sky in Gauss is
\begin{equation}
   B_{\perp}^{\mathrm{DCF}}=\xi\sqrt{4\pi\rho}\frac{\sigma_\nu}{\sigma_\psi}
   \label{eq:B-DCF}
,\end{equation}
where we assume that the magnetic field is frozen in the medium and
the dispersion of $\sigma_{\psi}$ is due to transverse and
incompressible Alfv\'en waves. In Eq.~(\ref{eq:B-DCF}),
$\rho=\mu n m_H$ is in g cm$^{-3}$, $\sigma_{\nu}$ is cm s$^{-1}$ and
$\sigma_{\psi}$ is in rad. The correction factor $\xi$ is usually
taken as $0.3\le\xi\le0.5$ and it has been derived by simulations of
MHD turbulence in molecular clouds \citep{Ostriker2001, Padoan2001,
  Heitsch2001, Falceta-Goncalves2008}. To avoid an overestimate of
$B_{\perp}^\mathrm{DCF}$, $\xi$ must be applied in case of strong
magnetic field (i.e.\ $\sigma_{\psi}\le25^{\circ}$).
\citet{Crutcher2003} however pointed out that self-gravitating cores
were not properly resolved in those simulations, since the simulations
were halted after the formation of dense filaments because of
insufficient resolution to follow the evolution further
\citep{Crutcher2003}.  Moreover, \citet{Cho2016} argued that in the
presence of averaging (e.g.\ on $\sigma_\psi$ along the line of sight
or on the polarisation angle within the telescope beam) the DCF
method tends to overestimate the magnetic field strength and thus they
proposed a correction factor $0.7\le \xi\le1.0$. Therefore,
considering all these effects, we decided to use a correction factor
$\xi=0.5$, as proposed by \citet{Crutcher2003}.  We also assume a mean molecular weight of
$2.8 m_H$ and we use an average number density $n=10^6$ cm$^{-3}$ for
denser cores MM4 and MM8,  while $n=10^5$ cm$^{-3}$ for
less dense cores MM3 ($n$ are from L17).

The angular dispersion of the
local magnetic field orientations has been estimated over the selected
pixels in each POLI images, considering
\begin{equation}
\sigma_\psi =\sqrt{\langle(\Delta\psi)^2\rangle}
\label{eq:sigma-psi}
\end{equation}
\begin{equation}
\Delta\psi = 0.5 \times \mathrm{arctan} \left(\frac {Q\langle U\rangle - \langle Q\rangle U}{Q\langle Q \rangle - \langle U\rangle U}\right)
\label{eq:delta-psi}
,\end{equation}
where $\psi$ is the polarisation orientation angle.

The L17 authors observed some molecular species emitted from the same
cores that we observed in our data. We make use of CH$_3$OH lines from L17 (either from their Table 2 or
their spectra) to estimate the velocity dispersion
$\sigma_{\nu}\sim \Delta\nu / \sqrt{8\ln 2}$.  Then we calculate the
$B_{\perp}^\mathrm{DCF}$ for the cores MM3, MM4, MM8, whose values are
reported in Tab.~\ref{tab:Bpar}.

The formal uncertainties on the magnetic field strength are computed
considering the errors propagation on the angle and velocity
dispersion. However the absolute uncertainties for the DCF method are
larger because of the uncertainties on the density $n$, which
we assume to be in the range $n=10^5 \sim 10^6$ cm$^{-3}$ ,  considering the error
on the correction factor of the order of 30\%
\citep{Crutcher2003}.

The estimates of $\sigma_{\psi}$ were performed using the
``classical'' variance, calculated around the circular
mean \citep{jammalamadaka2001,pewsey2013}, selecting an individual
region for MM3, MM4, and MM8.  MM4 presents a central core and it is
elongated to the west.  The L17 work does not report any velocity dispersion
for MM3, therefore we assumed the same values of velocity dispersion
used for
MM4. 
MM8 presents also a complex structure that has two filaments of gas
with the shape of spiral arms around a central core.  MM7, MM8c, MM9,
and MM11a are very compact sources, showing no specific structure.
They are too close to a point source to obtain reasonable estimates of
the magnetic field using them alone. Therefore, we estimated
$\sigma_{\psi}$ considering only the entire core MM8.

However, the magnetic field values obtained from this method should be
considered as order-of-magnitude estimates because they do not fully
encompass the complexity of the field dynamics in each part of the
protostar. For example, the value of $\sigma_\psi$ is an average of
magnetic field vectors and it is really difficult to determine the
region in which this average must be computed, especially in sources
showing complicated structures such as G9.62. In addition it is not clear
which turbulent velocity is relevant for the dust (we used CH$_3$OH in
Tab.~\ref{tab:Bpar}) and this results in one more uncertainty in the
DCF method.

\begin{figure*}
\centering
\includegraphics[width=.7\columnwidth,bb=83 35 576 730,clip, angle=-90]{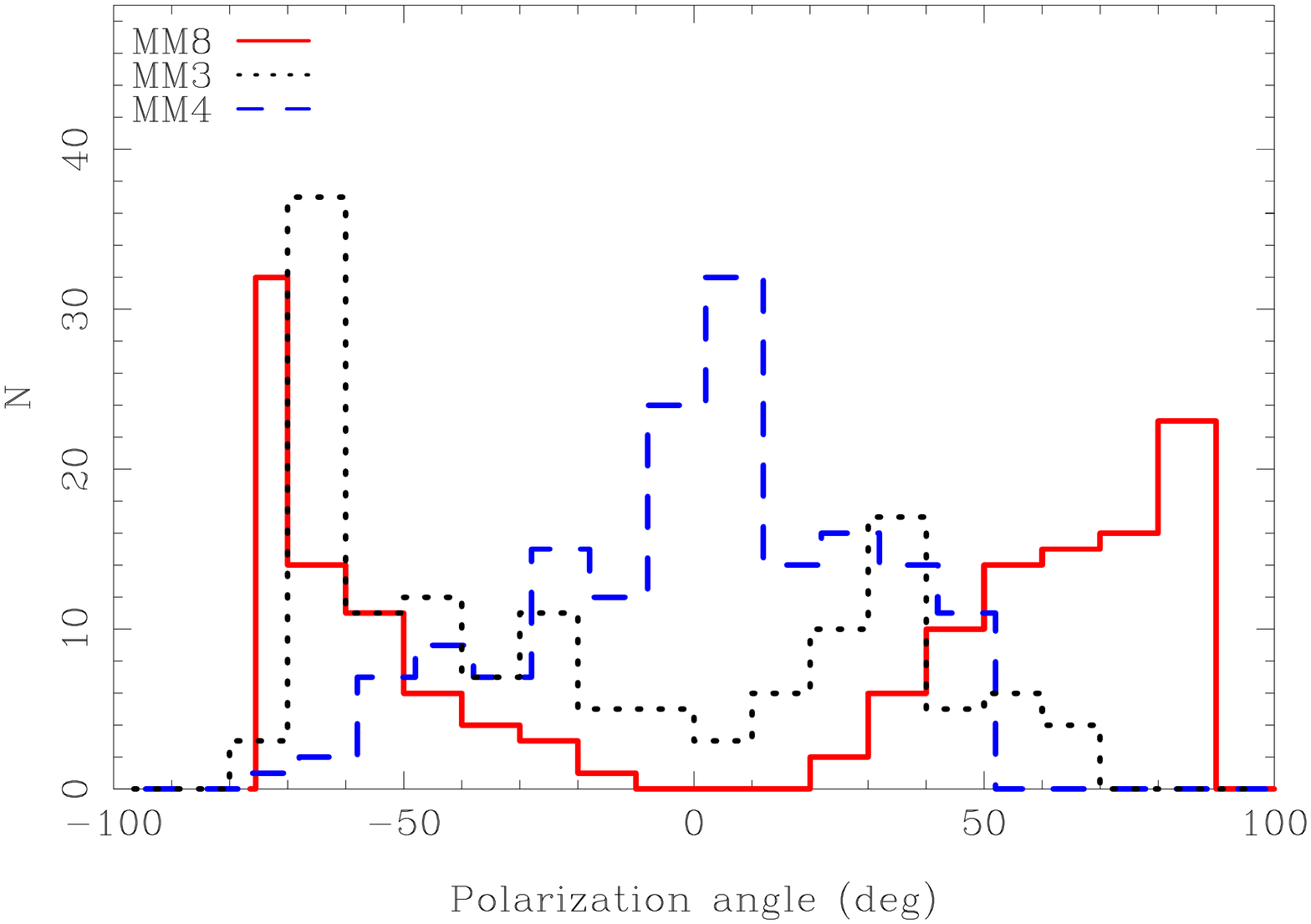}
\includegraphics[width=.7\columnwidth,bb=83 35 576 730,clip, angle=-90]{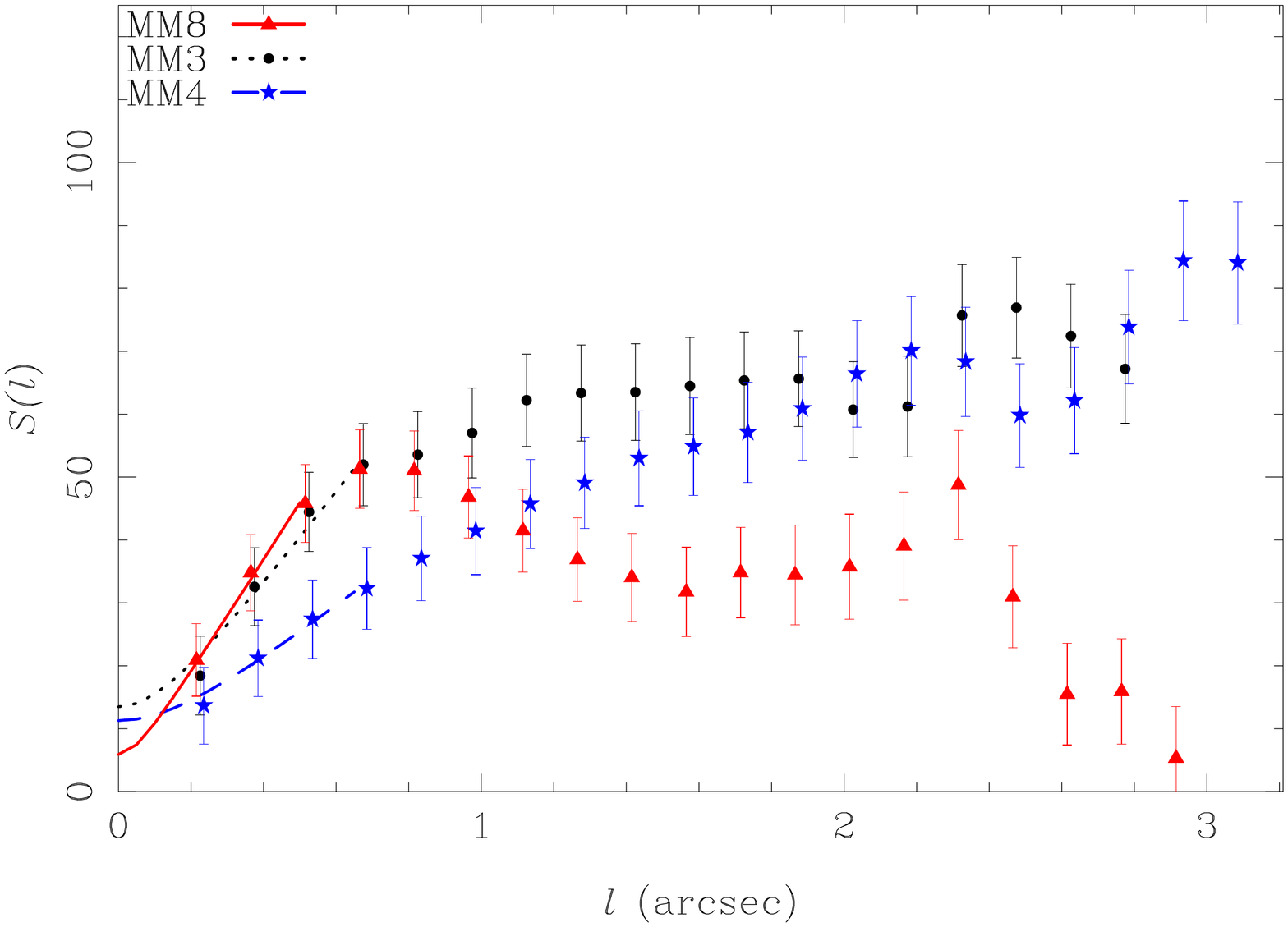}
\caption{Left panel: histograms of polarisation orientation angle
  $\psi$, towards the sources MM3, MM4, and MM8. Nyquist sampling was applied and each bin is
  10$^{\circ}$ wide.  Right panel: square root of the SF of the
  polarisation orientation angles, $S(l)$ for the cores MM3, MM4,
  and MM8 represented by the black dots, blue stars, and red
  triangles, respectively. The best fits for MM3, MM4, and MM8 are
  plotted as the dotted black line, dashed blue line, and solid red
  line, respectively. In the SF
  plot, we applied Nyquist sampling and a bin of 0.15$\arcsec$. }
\label{Fig:hisogram}
\end{figure*}

Following \citet{Planck_coll2016}, in Fig.~\ref{Fig:hisogram}
  we plot the histogram of the distribution of polarisation
  orientation angles $\psi$ towards the regions of MM3, MM4, and MM8.
Each bin is 10$^{\circ}$ wide.  From the plot, all the cores present a
large distribution of polarisation orientation angles. A broad
distribution of polarisation orientation angles can imply an
overestimate of $\sigma_\psi$ and consequently it can cause an
underestimate of the magnetic field. For example, MM8 exhibits a
spiral-shaped pattern of the polarisation segments resulting in a
histogram with a large distribution of $\psi$ and its magnetic field
is probably underestimated.

Moreover, the magnetic field structure inside a dense molecular cloud
may be subject to effects that are not considered in the DCF analysis,
such as differential rotation, gravitational collapse, or expanding
HII regions. Therefore, we may observe a distortion in polarisation
position angles due to large-scale non-turbulent effects. As a
consequence, the dispersion values measured about mean fields, assumed
to be straight, can be much larger than should be attributed to MHD waves or
turbulence \citep{Hildebrand2009, Houde2009}.

\subsubsection{Structure function of the polarisation orientation angles}
\label{sec:structure_function}
A possible way to overcome the problem of the underestimate of $B_{\perp}^\mathrm{DCF}$
is by measuring the structure function (SF), also known
as dispersion function of the polarisation orientation angles
\citep{Hildebrand2009, Houde2009, Koch2010}. It consists in computing the mean
absolute difference between polarisation angles as a function of their
displacement $l$.
Large values of the SF indicate large variations,
while small values express a small dispersion between measured
polarisation angles. The advantages of this method are that the
SF does not depend on any model of the large-scale
field. Nevertheless, it can only be applied under the same
conditions as the DCF method, i.e.\ a smooth low-noise polarisation
image, well-known densities, and moderately uniform gas velocities.  This
method, however, provides an estimate of the turbulence to
large-scale magnetic field strength ratio.
Following \citet{Hildebrand2009}, \citet{Houde2009}, and \citet{Koch2010}, in
the right panel of Fig.~\ref{Fig:hisogram} we plot the magnetic field
dispersion $S(l)$, which is the square root of the SF of the
polarisation orientation angles towards each source
$S^2(l)=b^2+m^2l^2+\sigma^2_M(l)$, where $b$ represents the turbulent
contribution to the angular dispersion, $m^2$ is a constant, and
$\sigma^2_M(l)$ is the uncertainty on the polarisation
angles.  We consider a displacement $l> 0.15\arcsec$ to avoid length
scales smaller than the beam. In the plot, the data are binned
with a width of 0.15$\arcsec$ and using Nyquist
sampling.
The errors are computed as the standard deviation of the values inside
the bin.
The magnetic field is composed by a large-structured field $B_0$ and
by a random component $B_t$ and their relationship is given by

\begin{equation}
  \frac{B_t}{B_0}=\frac{b}{\sqrt{(2-b^2)}} \quad.
    \label{eq:Bt/B0}
\end{equation}
 
Considering the same assumptions valid for the DCF method,
i.e.\ incompressible and isotropic turbulence, magnetic field
frozen into the gas, and dispersion of the $B_{\perp}$ orientation
originating in transverse incompressible Alfv\'en waves,

\begin{equation}
  \frac{B_t}{B_0}=\frac{\sigma_\nu}{\sigma_A} \quad,
  \label{eq:Bt/B0-Alfven}
\end{equation}
where $\sigma_A=B_0(4\pi\rho)^{-\frac{1}{2}}$ is the Alfv\'en velocity.

Combining Eq.~(\ref{eq:Bt/B0}) and Eq.~(\ref{eq:Bt/B0-Alfven}) we
obtain the magnetic field component on the plane of the sky computed
using the SF method

\begin{equation}
   B_{\perp}^{\mathrm{SF}}=\sqrt{4\pi\rho}\frac{\sigma_\nu\sqrt{2-b^2}}{b}  \quad.
   \label{eq:B-SF_originale}
\end{equation}

When $B_t \ll B_0$, then Eq.~(\ref{eq:B-SF_originale}) can be approximated
\begin{equation}
   B_{\perp}^{\mathrm{SF}}\simeq\sqrt{8\pi\rho}\frac{\sigma_\nu}{b} \quad.
   \label{eq:B-SF}
\end{equation}
 
For the details of the mathematical demonstration of
  the previous formulae, we refer to Appendix A in \citet{Hildebrand2009} and
  Appendix D in \citet{Planck_coll2016}.  By fitting $S^2(l)$, we can
  derive the intercept of the fit $b^2$, and its square root gives an
  alternative measure for the dispersion of polarisation angles
  $\sigma_\psi$ (see Tab.~\ref{tab:Bpar}).  In Tab.~\ref{tab:Bpar} we
  also report the magnetic field component on the plane of the sky
  computed using the SF method and the ratio $B_t/B_0$ showing that
  the large-scale magnetic field dominates the turbulent
  component. We considered the regions MM3, MM4, and MM8.
As described in \citet{Houde2009} and \citet{Koch2010}, we estimated the
number of turbulent cells N$\sim$2 from 

\begin{equation}
   N=\frac{(\delta^2+2W^2)\Delta'}{\sqrt{2\pi}\delta^3} 
   \label{eq:N_turb}
,\end{equation}

where $W$ is beam radius, $\Delta'$ is the effective depth of the
molecular cloud along the line of sight, and $\delta$ is the turbulent
correlation length.  Considering the distance of our source, our beam
radius ($\sim0.15\arcsec$) results in a physical size of $W\sim3.7$
mpc.  We assumed a $\Delta'\sim40$ mpc, which is approximately the size
of our cores and a $\delta\sim10$ mpc, which is comparable to
estimates made for other similar cores \citep{Girart2013, Frau2014,
  Juarez2017}.  Therefore, the beam effect correction results
$\sim\sqrt{N}=1.4$, which does not produce a significant change in our
$B_t/B_0$ ratio. This is consistent with the view proposed by
\citet{Koch2010}, where the beam correction is important for low-resolution data, but it is less important for high-resolution
observations.

\subsubsection{Turbulent-to-magnetic field energy ratio}
\label{sec:turbulence/B}

We can determine the one-dimensional Alfv\'en velocity
$\sigma_A\sim 15$ km s$^{-1}$, assuming
that the component along the line of sight is the same on the plane of
the sky and considering an approximate mean magnetic field strength in
the plane of the sky of the order of 11 mG (assumed from B$_\perp^\mathrm{SF}$ values in
Tab.~\ref{tab:Bpar}) and a typical number density of 10$^6$ cm$^{-3}$
(L17).

The comparison of $\sigma_A$ with the one-dimensional velocity
dispersion for each individual core $\sigma_\nu$
(Tab.~\ref{tab:Bpar}) indicates that any turbulent or infall velocity
is below the Alfv\'enic velocity.  The ratio
  $\sigma_\nu/\sigma_A \sim 0.16$ is consistent with the values that
  we find for $B_t/B_0$ ratio. We estimate the ratio of
turbulent-to-magnetic energy as $\gamma \sim 3(\sigma_\nu/\sigma_A)^2$
\citep{Beuther2018, Girart2009}. Using the larger line-of-sight
velocity dispersion found for the cores in MM8 ($\sigma_\nu=2.35$ km
s$^{-1}$), the turbulent-to-magnetic energy ratio is
$\gamma \sim 0.07$.  Thus, the magnetic energy along the filament
appears to dominate over the turbulent energy indicating that magnetic
field seems to be playing a more dominant role than turbulence.

The important role played by the magnetic field might also influence the morphology and fragmentation of the filament, resulting in a
tiny number of low-mass protostars. Simulations performed by
\citet{Inoue2013} suggested that because of enhanced magnetosonic
speed and turbulence behind the shock, the effective Jeans mass and
the mass accretion rate can have a larger value in this region,
triggering the formation of mainly massive cores. From their simulations,
which include MHD contributions, massive star formation is
naturally triggered when the massive filament collapses globally.
However, the lack of low-mass protostars along the filament could also
be justified considering other theories such as the ``collect and
collapse'' process, as discussed in L17. External heating coming from
the two \Hii regions (B and C see Fig.~\ref{Fig:Spitzer}) and internal
heating generated by outflows from hot molecular cores (such as MM8 or
MM4) increased the thermal Jeans masses and consequently suppressed
the fragmentation.

\subsection{Mass to magnetic flux ratio}
\label{sec:mass-to-flux}

Another way to quantify if the magnetic field can affect the
fragmentation and formation of new stars is to estimate the
mass-to-flux ratio $(M/\Phi_B)$. This ratio defines the stability of a region
and whether a static magnetic field can support a cloud against
gravitational collapse \citep{Crutcher1999, Troland2008}. The
mass-to-flux ratio in units of critical mass-to-flux ratio is
\begin{equation}
\label{eq:Mtoflux}
\lambda =(M/\Phi_B)_{obs}/(M/\Phi_B)_{crit}\sim 7.6 \times 10^{-24} N_{H_2}/B   \quad,
\end{equation}  

where $(M/\Phi_B)_{crit}$ and $(M/\Phi_B)_{obs}$ are the
theoretically determined critical mass-to-flux ratio
\citep{Nakano1978} and the observed mass-to-flux ratio, respectively, with $N_{H_2}$
in cm$^{-2}$ and $B$ in mG.

Using the column densities and the average magnetic field obtained
from the SF analysis reported in
Tab.~\ref{tab:continuum_sources} and Tab.~\ref{tab:Bpar}, respectively,
we get $\lambda$ varying roughly from 8 to 13. Looking at the mass-to-magnetic-flux ratio, the
G9.62 clump appears to be super-critical, which is consistent with
\citet{Cortes2006} because $\lambda$ is several order of magnitude above the
critical value and indeed star formation is occurring in the filament.

However, the column density used to determine $\lambda$ naturally
implies an average density that is much higher than the density adopted in the
magnetic field calculation (Eqs.~\ref{eq:B-DCF}, \ref{eq:B-SF}), when assuming a
source size along the line of sight similar to that on the plane of
the sky. Hence, the magnetic field strength measured is not
necessarily representative of the magnetic field supporting the core,
which means that $\lambda$ is strictly an upper limit and could be
more than an order of magnitude smaller. This illustrates one of the
main uncertainties of estimating $\lambda$ from dust polarisation
measurements. In addition, Eq.~(\ref{eq:Mtoflux}) comes from cloud
models computed from initially uniform, spherical clouds with
initially uniform magnetic fields. If the mass is differently
funnelled along magnetic flux lines, then it results in a
different $(M/\Phi_B)_{crit}$, which can be higher by a factor of 2
\citep{Mouschovias1991}. Moreover, if the cores are supported
primarily by static magnetic fields, they should have flattened
structures and therefore $N_{H_2}$ should be measured parallel to the
magnetic field to correctly estimate the mass-to-flux ratio.

Thus, taking into account all the approximations and uncertainties on
$\lambda$, on the column densities and on the magnetic field strength
obtained with the SF method, these values of the mass-to-magnetic-flux
ratio should be considered with caution.

\subsection{Detection of thermal line polarisation}
\label{sec:linepol}

In one of the spectral windows, molecular emission was detected from
the core MM4a with a peak at 337.6 GHz. We performed the continuum
subtraction and we detected linearly polarised emission only from the
core MM4a and in few channels around the peak of the line
emission. Therefore we exclude any dependence on the position within
one-third of the primary beam. In our observations we do not have enough
spectral resolution to identify unambiguously the molecular
species. Considering for MM4a a velocity of $\sim$2 km s$^{-1}$ (from L17),
methanol or sulphur dioxide are the most likely candidates, since they
are species that have already been observed in the core (L17) and both of these
present several lines at the frequency of the observed peak.

\begin{figure*}
 \centering
  \includegraphics[width=\columnwidth, bb=128 60 700 484, clip]{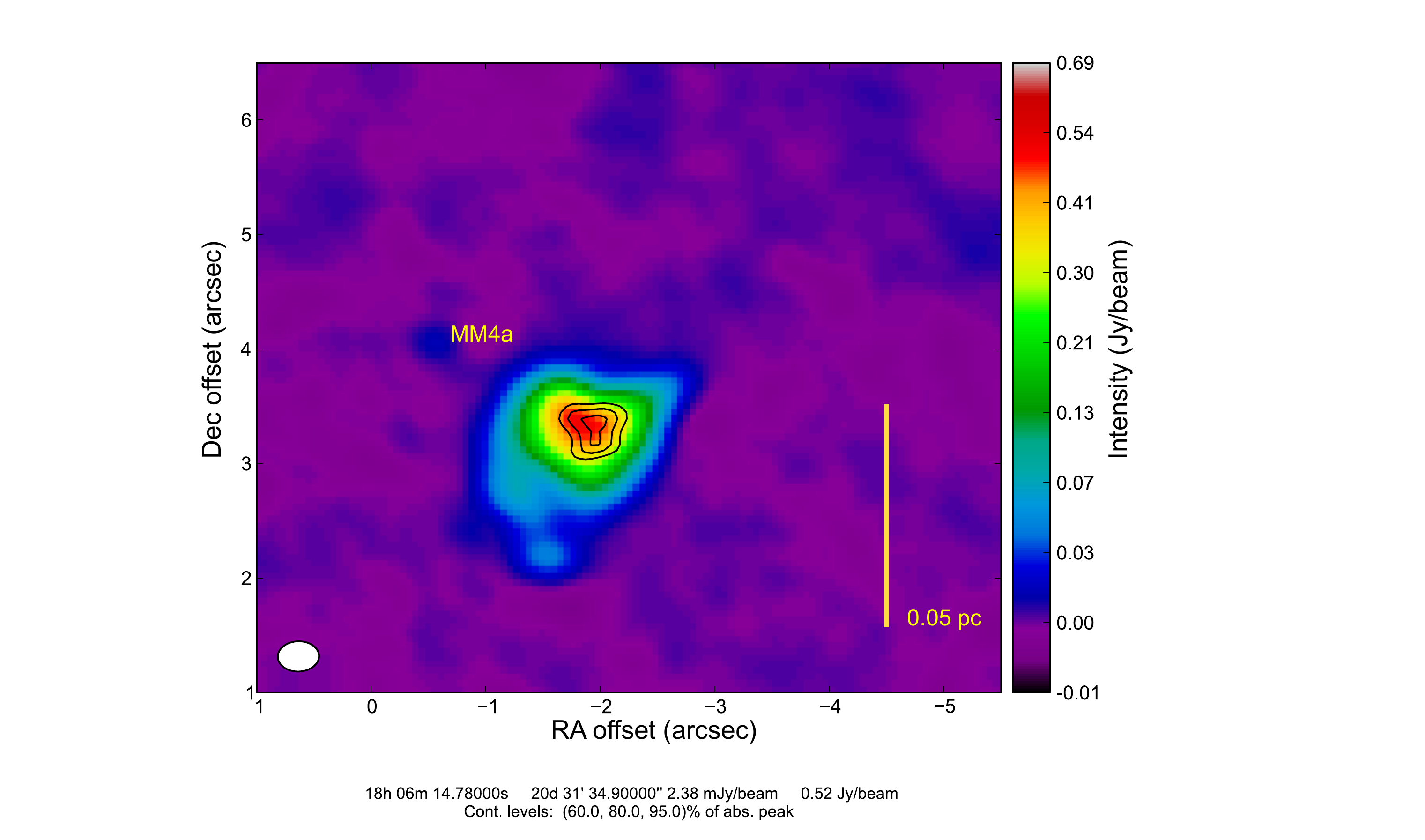}
  \includegraphics[width=\columnwidth, bb=128 60 700 484, clip]{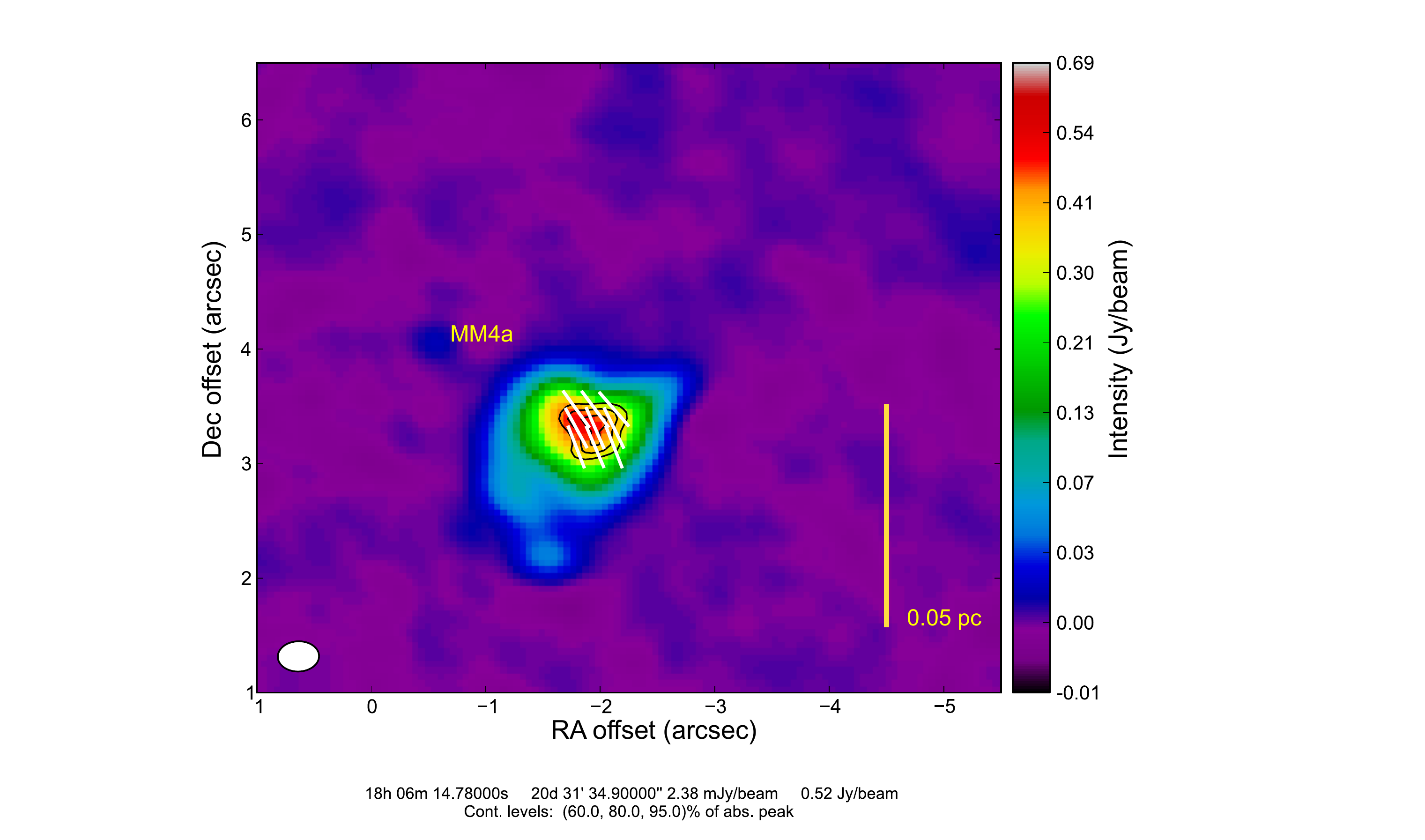} 
  \caption{Total intensity image of the thermal line emitted in the core MM4a. 
    Left panel: the contours represent linearly polarised emission and
    the levels are (1.4, 1.9, 2.3) mJy beam$^{-1}$.  The linear
    polarisation peak intensity is 2.38 mJy beam$^{-1}$.  Right panel:
    the segments indicate linear polarisation vectors already rotated
    by 90$^{\circ}$ to show the orientation of the magnetic field. The
    sampling of the vectors are every 165 by 165 
    mas. In both panels, the offsets are relative to the same
    position as in Fig.~\ref{Fig:mosaico1} and the colour scale
    indicates the total intensity in Jy beam $^{-1}$, going from -0.01
    to 0.52 Jy beam $^{-1}$; the yellow bar indicates the physical scale of 0.05
    pc at the distance of the source.}
 \label{Fig:lineapol}
\end{figure*}

In Fig.~\ref{Fig:lineapol} we plot the total line intensity, linear polarisation contours, and vectors indicating the
polarisation direction already rotated of 90$^{\circ}$.
It is ambiguous, for the polarisation resulting from the
Goldreich-Kylafis effect, whether the magnetic field orientation is
identical to the polarisation orientation or whether a 90$^{\circ}$
rotation is necessary. We decided to plot the polarisation vectors
already rotated by 90$^{\circ}$. This decision was made under a few
assumptions based on the analysis performed by \citet{Cortes2005} for
the CO line.  The linear polarisation can be perpendicular to the
magnetic field lines in presence of a velocity gradient which is
smaller in the direction parallel to the magnetic field. If we
consider the direction of the magnetic field probed by our dust
observations and we look at the direction of the velocity gradient in
Fig.5 in L17, we see that this condition is valid for MM4. We are then
assuming an anisotropy of the radiation field which can allow an
emission polarised in the direction perpendicular to the magnetic
field. However, we emphasise that we are conscious of the ambiguous
behaviour of the polarisation resulting from the Goldreich-Kylafis
effect and that the work by \citet{Cortes2005} was proposed for
another molecule. Even if we do not have firm information about the
radiation field, we present the direction already rotated by
90$^{\circ}$ to show a tentative comparison between the three
independent methods to infer the magnetic field (see
Sect.~\ref{sec:paragone}). For a more complete analysis, more
detailed knowledge about the radiation field and the optical depth of
the molecule is needed.

The total intensity peak is 0.52 Jy beam
$^{-1}$, and the linear polarised intensity is 2.38 mJy beam
$^{-1}$. The polarisation threshold is 1.4 mJy beam $^{-1}$ and the rms
is 0.4 mJy beam $^{-1}$. The direction of the linear polarisation is
consistent with that of core MM4a observed with the dust, while it is
perpendicular to the direction of the vectors marking the elongated structure in
north-east direction. Because of the lack of spectral resolution and the
uncertainty in the identification of the molecule, we do not further
analyse the 336.7 GHz line.

\section{Discussion}
\label{sec:discussion}

\subsection{Magnetic fields in different evolutionary stages }
\label{sec:B_Evolution}

Since complex organic molecules are good tracers of core evolution
\citep{Qin2010}, L17 suggested that the presence of molecules such as
CH$_3$OH and CH$_3$OCHO in MM4 and MM8 (and relative subcores)
indicates that both cores are in a hot molecular core phase\footnote{
  \citet{Liu2017} performed SED fittings and obtained archived
  data from Herschel (70, 160, 250, 350, and 500 $\mu$m ), JCMT/SCUBA
  (450 and 850 $\mu$m continuum), APEX/LABOCA (875 $\mu$m continuum
  images), and SEST/SIMBA (1.2 mm continuum image). For the spectral
  lines analysis they made use of ALMA band 6 data. All the details
  are reported in Appendix A in L17. For further references see also
  Sec.2.}.  Moreover, again according to L17, MM8 is less evolved than
MM4 because MM8 is driving energetic outflows and it might be still
in accretion phase. MM3 is instead a starless core, probably at an
early evolutionary phase (L17).

Therefore the evolutionary sequence from less evolved core to
more evolved core is MM3 $\rightarrow$ MM8 $\rightarrow$ MM4.
Depending on the exact evolutionary stage, we could expect to observe
highly ordered magnetic fields in the more quiescent cores, while the
field in the more evolved protostars is less uniform.
As shown in Fig.~\ref{Fig:mosaicozoomN} and
Fig.~\ref{Fig:mosaicozoomS}, our band 7 observations reveal an
elongated structure for MM3, MM4, and MM8 and several subcores showing
polarised emission.

In MM3 we detected the strongest linear polarisation fraction observed
in G9.62 and the field lines are organised in a orderly
pattern.  However in MM3 the resulting $B_{\perp}$ is probably
  underestimated since, to perform the magnetic field analysis, we
  assumed the same emission of CH$_3$OH observed in MM4 and L17 did
  not report any line for this core. Moreover, although MM3 shows a
highly polarised flux, its spectral index is not really different from
the other cores.  Therefore, it may be possible that MM3 is at an
early evolutionary phase and the collapse did not start yet because
gravity was unable to overcome magnetic pressure in this
core.

MM4 is the most evolved core (L17 and references therein), since it
shows a strong continuum emission and no molecular outflow. It presents
a elongated profile in the north-west direction and the linear
polarisation vectors show a orderly patter.
MM4 shows a broad
distribution of $\psi$ (Fig.~\ref{Fig:hisogram}) but only one peak,
indicating the presence of one component and a less complex magnetic
field.
From Fig. 5 in L17, we can also see that MM4
presents a range of velocities from $\sim$1 km s$^{-1}$  to $\sim$3
km s$^{-1}$.

MM8 is in an early hot core stage showing outflows and clear
signatures of accretion (L17). In our total intensity image, MM8 exhibits
a structure along the north-south direction and the polarisation vectors
are organised in a peculiar spiral shape. This could be a signature of
compressed magnetic field due to infalling material that is
accreted by the young protostars, and the magnetic field lines are
still frozen in the medium. The broad distribution of polarisation orientation angles that we
find in Fig.~\ref{Fig:hisogram} could be due to the presence of separate components.
From Fig. 5 in L17, we can also see that MM8
 presents a range of velocities broader than MM4, spanning from $\sim$1 km s$^{-1}$ to $\sim$6
km s$^{-1}$.

Since we do not have molecular lines observations at the same
resolution in the same region in our Band 7 data, we could not perform any velocity field
analysis.
We decided to exclude MM3 from the comparison, since its
  magnetic field could be underestimated.  Thus, we focussed only on
MM8 and MM4; interpreting the magnetic field data in terms of the
evolutionary stages, we observe a magnetic field stronger in the
youngest core MM8 than in the more evolved source MM4. This is
consistent with the scenario in which magnetic fields are influencing
star formation and it is symptomatic of an evolution of the magnetic
field strength and morphology during the entire star formation
process.

At small scales of 0.1 pc, all the cores
in MM3, MM4, and MM8 present magnetic fields that are organised and
non-chaotic. Comparing the direction of the outflows studied by L17 with the
direction of the magnetic field indicated by our linear polarisation
observations, we see that they are perpendicular, which is partially
in disagreement with the finding of
\citet{Hull2013}. Since there is a correlation between the orientation
of the axes of the molecular outflow and the magnetic field, it could
be a signature of a magnetically regulated evolution.  The magnetic
field could be strong enough to funnel the gas along the field lines
and regulate the accretion and disc orientation.

Furthermore, it seems that the cores presenting a uniform magnetic
field, such as MM3 or MM4, are less fragmented than the other cores
not showing polarisation. We also estimate on average a
turbulent-to-magnetic energy ratio of $\gamma\sim$0.07, indicating
that the magnetic energy along the filament dominates over the
turbulent energy.
Moreover the orientations of the magnetic field segments in
our ALMA observations (core-scale) are consistent with those in JCMT/POL-2
observations (clump-scale) \citep{Liu2018}.
Therefore, this can be a further evidence that magnetic fields have a
strong influence on the star formation processes in this clump,
playing an important role in the collapse and fragmentation phase.

\subsection{Tentative comparison between dust continuum, thermal line, and maser data}
\label{sec:paragone}

\begin{figure*}
 \centering
  \includegraphics[width=0.8\textwidth, bb=0 18 355 468, clip]{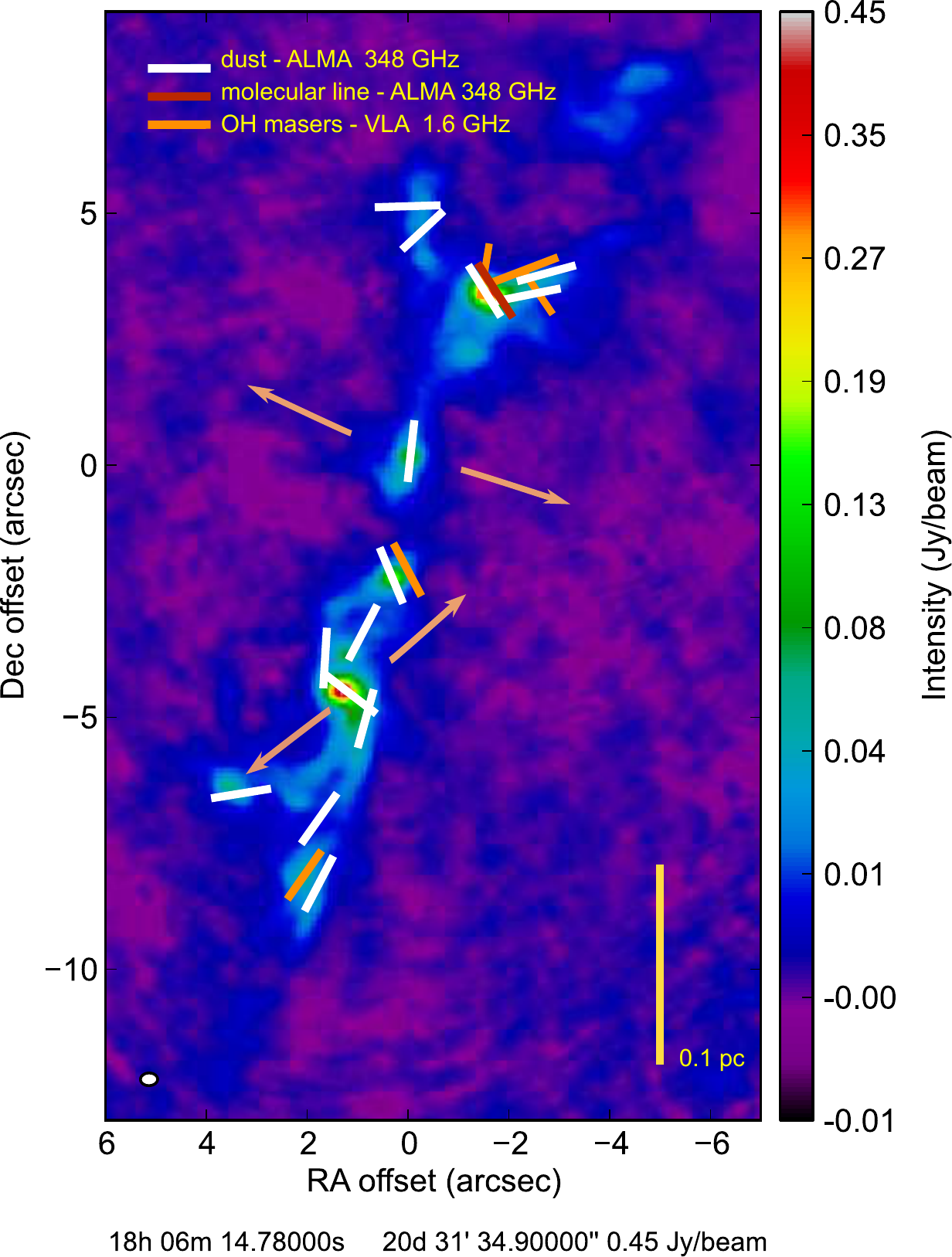}
  \caption{Total intensity image of the star forming region G9.62+0.19
    and linear polarisation segments, rotated by 90$^\circ$,
    indicating the average direction of the magnetic field on scales of
    $\sim 0.5\arcsec \times 0.5\arcsec$. The different colours in the
    segments delineate the direction of the magnetic field as detected by
    different data at different wavelengths. The white and brown bars
    indicate dust observations and molecular thermal line
,    respectively, from our ALMA band 7 polarised data. The orange bars
    indicate OH maser linearly polarised observations
    \citep{Fish2005maser,Fish2006}. Arrows denote the direction of the
    outflows (L17). The offsets are relative to the absolute position
    in Fig.~\ref{Fig:mosaico1}. The white ellipse represents the
    beam. The colour scale goes from -0.01 to 0.45 Jy beam $^{-1}$ and
    the yellow bar indicates the physical scale of 0.1 pc, at the
    distance of the source.}
 \label{Fig:summary}
\end{figure*}

In this section we want to discuss a tentative comparison of the
resulting magnetic field for the G9.62 clump obtained by three
different and independent methods: the analysis of the linear
polarised light emitted by dust in the continuum (DCF and SF methods),
the analysis of the linear polarised light emitted by thermal
molecular transitions, and the analysis of the polarised maser
emission.

In Fig.~\ref{Fig:summary}, we plot a summary of the magnetic field
directions obtained from our ALMA 348 GHz data and from 1.6 GHz OH
masers observations \citep{Fish2005maser,Fish2006}. In this figure,
the linear polarisation segments are overplotted on the total
intensity image of the filament. The segments are already rotated by 90$^\circ$, indicating the direction of the magnetic field as inferred
by different observations at different wavelengths. In general, the
magnetic field exhibits an ordered structure along the direction of
this filament. Comparing the direction of the outflows studied by L17
with the direction of the magnetic field denoted by our linear
polarisation observations, we see that they are perpendicular. In
addition, if we assume a rotation of 90$^\circ$, the linear
polarisation vectors obtained from the thermal line analysis in the
core MM4a point the same direction indicated by the polarised dust
(Fig.~\ref{Fig:summary}).  Moreover, the averaged magnetic field
strength on the plane of the sky obtained from our analysis (of the
order of 11mG) is comparable with the component along the line of
sight, already proposed by OH and CH$_3$OH masers observations
\citep{Vlemmings2008}. Indeed OH maser Zeeman splitting observations
indicate a magnetic field strength of $\sim$5 mG. The magnetic field
determined from the 6.7 GHz methanol masers also revealed a similar
strength.  The linear polarisation vectors from the OH masers appear
consistent with our linearly polarised data, but unfortunately the
small number of maser observations hinders a quantitative
demonstration. Even though the thermal line, masers, and dust
polarisation probe very different physical processes and likely
very different volumes as well, observations towards the massive protostar
IRAS18089-1732 \citep{Daria2017_IRAS18089, Beuther2010} showed that
the same magnetic field can affect both the small scale of few
astronomical units around the protostar (probed by masers) and the
large scale of the torus (probed by dust). The alignment of the
geometry between the dust emission (coming from the core) and the
thermal line emission (coming from the envelope) has also been found
in DR21(OH) \citep{Lai2003,Cortes2005}, suggesting a connection
between the magnetic field in the core and that in the envelope. So
further observations of polarised thermal line, masers, and dust are
needed to investigate the alignment of the magnetic field at different
scales of star forming regions.

\section{Conclusions}
\label{sec:conclusions}

In this paper, we presented our investigation of the magnetic field
morphology of the well-known high-mass star forming region G9.62+0.19. We
analysed ALMA band 7 continuum observations and we identified 23 cores
and substructures. For these we reported the position, peak flux
densities, integrated flux, deconvolved major and minor axes,
position angles, and spectral index.

We detected linear polarised emission from two cores and relative
subcores in the northern field (MM3 and MM4) and from four in the
southern field (MM7, MM8, MM9, MM11a). One of these, MM3a showed a
linear polarisation fraction of $\sim 10$\%. For these we studied the
magnetic field strength on the plane of the sky component $B_\perp$,
comparing the Davis-Chandrasekhar-Fermi (DCF) and the structure
function (SF) analysis. Through the SF we derived estimates of the
magnetic field strength along the line of sight that were larger than
the values obtained by the classical DCF method.

By comparison with band 6 observations \citep{Liu2017}, for some of
the cores we obtained the spectral index $\alpha$ and the spectral
index for the dust opacity $\beta$ and we estimated the N(H$_2$)
column densities and the masses. We found that the core masses range
roughly between 10 M$_\odot$  and 40 M$_\odot$ and the N(H$_2$) column densities
between 3 $\times$ 10$^{24}$ and 5 $\times$ 10$^{25}$ cm$^2$. Because of
the large uncertainties, mainly determined by the errors on the dust
model, on the spectral index and optical depth, the errors of
the masses and column densities are difficult to quantify and they
could be more than a factor of 5.

We proposed an evolutionary sequence for the magnetic field of the two
hot cores MM4 and MM8, indicating that MM4 is more evolved than MM8 because of the presence of a weaker magnetic field than that observed
in MM8. MM3 is a starless core at an early evolutionary phase and the
collapse did not start yet, because in this core gravity was probably
unable to overcome magnetic pressure yet; its magnetic field is indeed
probably underestimated since from previous observations we do not have
information about molecular lines coming from this core.

In general the magnetic field seemed to follow the direction of the
filament, and it was perpendicular to the direction of the outflows
emitted by some massive protostellar cores such as MM8a, MM7, and
MM6. The cores that presented polarisation appeared to be less
fragmented than those not showing polarised emission.
At scales less than 0.1 pc, the magnetic field showed a neat and
ordered pattern of polarisation vectors.

We also detected linearly polarised molecular line, thermally emitted
probably by methanol or carbon dioxide. Moreover, from the SF analysis we obtained an average magnetic field strength of
the order of 11 mG. The magnetic field strength on the plane of the
sky obtained from our analysis is comparable with the component along
the line of sight, already detected from previous OH and CH$_3$OH
masers observations on the core MM4b and MM7 \citep{Fish2005maser,
  Vlemmings2008}, suggesting that at scale of less of 0.05 pc the
magnetic field role could be important in this region.

We estimated the ratio of turbulent-to-magnetic energy and we found on
average a turbulent-to-magnetic energy ratio of $\gamma\sim$0.07,
indicating that the magnetic energy along the filament dominates over
the turbulent energy.

However, because of the uncertainties of the DCF and the SF methods,
which are inevitably reflected on the column densities and masses,
further investigations are needed to properly evaluate the magnetic
field strength in the G9.62 region. The use of combined observations
of masers, dust, and molecular lines  contribute towards
understanding the role of the magnetic field at different scales. As
demonstrated by our observations, ALMA has the capabilities to detect
the weak Goldreich-Kylafis effect \citep{Goldreich1982} occurring in
molecular lines. Therefore this instrument can add more details and constraints
on the magnetic field morphology and strength and can help to infer,
for example, more precise mass-to- magnetic-flux ratios.

\begin{acknowledgements}

  We thank the referee Patrick Koch for the precise comments and suggestions that
  have contributed to the improvement of this work.

  The research leading to these results has received funding from the
  European Research Council under the European Union's Seventh
  Framework Programme (FP/2007-2013) / ERC Grant Agreement
  n. 614264. This paper makes use of the following ALMA data:
  ADS/JAO.ALMA \#2015.1.01349.S. ALMA is a partnership of ESO
  (representing its member states), NSF (USA) and NINS (Japan),
  together with NRC (Canada) and NSC and ASIAA (Taiwan) and KASI
  (Republic of Korea), in cooperation with the Republic of Chile. The
  Joint ALMA Observatory is operated by ESO, AUI/NRAO and
  NAOJ. D.D. thanks the Nordic ALMA Regional Centre
  (ARC) node based at Onsala Space Observatory for support, in particular Tobia
  Carozzi and Ivan Mart\'i-Vidal. The Nordic ARC node is funded
  through Swedish Research Council grant No 2017-00648. D.D also
  thanks Akimasa Kataoka and Kei Tanaka for interesting
  discussions. H.B. acknowledges support from the European Research
  Council under the Horizon 2020 Framework Program via the ERC
  Consolidator Grant CSF-648505. JMG and JMT are supported by the
  MINECO (Spain) AYA2017- 84390-C2 coordinated grant.
\end{acknowledgements}

\bibliographystyle{aa}
\bibliography{../bibliografia}

\begin{thebibliography}{97}
\expandafter\ifx\csname natexlab\endcsname\relax\def\natexlab#1{#1}\fi

\bibitem[{{Alves} {et~al.}(2018){Alves}, {Girart}, {Padovani}, {Galli},
  {Franco}, {Caselli}, {Vlemmings}, {Zhang}, \& {Wiesemeyer}}]{Alves2018}
{Alves}, F.~O., {Girart}, J.~M., {Padovani}, M., {et~al.} 2018, \aap, 616, A56

\bibitem[{{Andersson} {et~al.}(2015){Andersson}, {Lazarian}, \&
  {Vaillancourt}}]{Andersson2015}
{Andersson}, B.-G., {Lazarian}, A., \& {Vaillancourt}, J.~E. 2015, \araa, 53,
  501

\bibitem[{{Attard} {et~al.}(2009){Attard}, {Houde}, {Novak}, {Li},
  {Vaillancourt}, {Dowell}, {Davidson}, \& {Shinnaga}}]{Attard2009}
{Attard}, M., {Houde}, M., {Novak}, G., {et~al.} 2009, \apj, 702, 1584

\bibitem[{{Banerjee} \& {Pudritz}(2007)}]{Banerjee2007}
{Banerjee}, R. \& {Pudritz}, R.~E. 2007, \apj, 660, 479

\bibitem[{{Beuther} {et~al.}(2018){Beuther}, {Soler}, {Vlemmings}, {Linz},
  {Henning}, {Kuiper}, {Rao}, {Smith}, {Sakai}, {Johnston}, {Walsh}, \&
  {Feng}}]{Beuther2018}
{Beuther}, H., {Soler}, J.~D., {Vlemmings}, W., {et~al.} 2018, \aap, 614, A64

\bibitem[{{Beuther} {et~al.}(2010){Beuther}, {Vlemmings}, {Rao}, \& {van der
  Tak}}]{Beuther2010}
{Beuther}, H., {Vlemmings}, W.~H.~T., {Rao}, R., \& {van der Tak}, F.~F.~S.
  2010, \apjl, 724, L113

\bibitem[{{Bonnell} \& {Bate}(2006)}]{Bonnell2006}
{Bonnell}, I.~A. \& {Bate}, M.~R. 2006, \mnras, 370, 488

\bibitem[{{Cesaroni} {et~al.}(1994){Cesaroni}, {Churchwell}, {Hofner},
  {Walmsley}, \& {Kurtz}}]{Cesaroni1994}
{Cesaroni}, R., {Churchwell}, E., {Hofner}, P., {Walmsley}, C.~M., \& {Kurtz},
  S. 1994, \aap, 288, 903

\bibitem[{{Chandrasekhar} \& {Fermi}(1953)}]{Chandrasekhar-Fermi1953}
{Chandrasekhar}, S. \& {Fermi}, E. 1953, \apj, 118, 113

\bibitem[{{Cho} \& {Lazarian}(2005)}]{Cho2005}
{Cho}, J. \& {Lazarian}, A. 2005, \apj, 631, 361

\bibitem[{{Cho} \& {Yoo}(2016)}]{Cho2016}
{Cho}, J. \& {Yoo}, H. 2016, \apj, 821, 21

\bibitem[{{Commer{\c c}on} {et~al.}(2011){Commer{\c c}on}, {Hennebelle}, \&
  {Henning}}]{Commercon2011}
{Commer{\c c}on}, B., {Hennebelle}, P., \& {Henning}, T. 2011, \apjl, 742, L9

\bibitem[{{Cortes} \& {Crutcher}(2006)}]{Cortes2006}
{Cortes}, P. \& {Crutcher}, R.~M. 2006, \apj, 639, 965

\bibitem[{{Cortes} {et~al.}(2005){Cortes}, {Crutcher}, \&
  {Watson}}]{Cortes2005}
{Cortes}, P.~C., {Crutcher}, R.~M., \& {Watson}, W.~D. 2005, \apj, 628, 780

\bibitem[{{Crutcher}(1999)}]{Crutcher1999}
{Crutcher}, R.~M. 1999, \apj, 520, 706

\bibitem[{{Crutcher}(2012)}]{Crutcher2012}
{Crutcher}, R.~M. 2012, \araa, 50, 29

\bibitem[{{Crutcher} {et~al.}(2003){Crutcher}, {Nutter}, \&
  {Ward-Thompson}}]{Crutcher2003}
{Crutcher}, R.~M., {Nutter}, D.~J., \& {Ward-Thompson}, D. 2003, in Bulletin of
  the American Astronomical Society, Vol.~35, American Astronomical Society
  Meeting Abstracts \#202, 772

\bibitem[{{Cudlip} {et~al.}(1982){Cudlip}, {Furniss}, {King}, \&
  {Jennings}}]{Cudlip1982}
{Cudlip}, W., {Furniss}, I., {King}, K.~J., \& {Jennings}, R.~E. 1982, \mnras,
  200, 1169

\bibitem[{{Dall'Olio} {et~al.}(2017){Dall'Olio}, {Vlemmings}, {Surcis},
  {Beuther}, {Lankhaar}, {Persson}, {Richards}, \&
  {Varenius}}]{Daria2017_IRAS18089}
{Dall'Olio}, D., {Vlemmings}, W.~H.~T., {Surcis}, G., {et~al.} 2017, \aap, 607,
  A111

\bibitem[{{Davis} \& {Greenstein}(1951)}]{Davis1951}
{Davis}, Jr., L. \& {Greenstein}, J.~L. 1951, \apj, 114, 206

\bibitem[{{Draine} \& {Weingartner}(1996)}]{Draine1996}
{Draine}, B.~T. \& {Weingartner}, J.~C. 1996, \apj, 470, 551

\bibitem[{{Draine} \& {Weingartner}(1997)}]{Draine1997}
{Draine}, B.~T. \& {Weingartner}, J.~C. 1997, \apj, 480, 633

\bibitem[{{Falceta-Gon{\c c}alves} {et~al.}(2008){Falceta-Gon{\c c}alves},
  {Lazarian}, \& {Kowal}}]{Falceta-Goncalves2008}
{Falceta-Gon{\c c}alves}, D., {Lazarian}, A., \& {Kowal}, G. 2008, \apj, 679,
  537

\bibitem[{{Fish} \& {Reid}(2006)}]{Fish2006}
{Fish}, V.~L. \& {Reid}, M.~J. 2006, \apjs, 164, 99

\bibitem[{{Fish} {et~al.}(2005{\natexlab{a}}){Fish}, {Reid}, {Argon}, \&
  {Zheng}}]{Fish2005maser}
{Fish}, V.~L., {Reid}, M.~J., {Argon}, A.~L., \& {Zheng}, X.-W.
  2005{\natexlab{a}}, \apjs, 160, 220

\bibitem[{{Fish} {et~al.}(2005{\natexlab{b}}){Fish}, {Reid}, \&
  {Menten}}]{Fish2005}
{Fish}, V.~L., {Reid}, M.~J., \& {Menten}, K.~M. 2005{\natexlab{b}}, \apj, 623,
  269

\bibitem[{{Frau} {et~al.}(2014){Frau}, {Girart}, {Zhang}, \& {Rao}}]{Frau2014}
{Frau}, P., {Girart}, J.~M., {Zhang}, Q., \& {Rao}, R. 2014, \aap, 567, A116

\bibitem[{{Garay} {et~al.}(1993){Garay}, {Rodriguez}, {Moran}, \&
  {Churchwell}}]{Garay1993}
{Garay}, G., {Rodriguez}, L.~F., {Moran}, J.~M., \& {Churchwell}, E. 1993,
  \apj, 418, 368

\bibitem[{{Girart} {et~al.}(2009){Girart}, {Beltr{\'a}n}, {Zhang}, {Rao}, \&
  {Estalella}}]{Girart2009}
{Girart}, J.~M., {Beltr{\'a}n}, M.~T., {Zhang}, Q., {Rao}, R., \& {Estalella},
  R. 2009, Science, 324, 1408

\bibitem[{{Girart} {et~al.}(2018){Girart}, {Fern{\'a}ndez-L{\'o}pez}, {Li},
  {Yang}, {Estalella}, {Anglada}, {{\'A}{\~n}ez-L{\'o}pez}, {Busquet},
  {Carrasco-Gonz{\'a}lez}, {Curiel}, {Galvan-Madrid}, {G{\'o}mez}, {de
  Gregorio-Monsalvo}, {Jim{\'e}nez-Serra}, {Krasnopolsky}, {Mart{\'{\i}}},
  {Osorio}, {Padovani}, {Rao}, {Rodr{\'{\i}}guez}, \& {Torrelles}}]{Girart2018}
{Girart}, J.~M., {Fern{\'a}ndez-L{\'o}pez}, M., {Li}, Z.-Y., {et~al.} 2018,
  \apjl, 856, L27

\bibitem[{{Girart} {et~al.}(2013){Girart}, {Frau}, {Zhang}, {Koch}, {Qiu},
  {Tang}, {Lai}, \& {Ho}}]{Girart2013}
{Girart}, J.~M., {Frau}, P., {Zhang}, Q., {et~al.} 2013, \apj, 772, 69

\bibitem[{{Girart} {et~al.}(2006){Girart}, {Rao}, \& {Marrone}}]{Girart2006}
{Girart}, J.~M., {Rao}, R., \& {Marrone}, D.~P. 2006, Science, 313, 812

\bibitem[{{Goldreich} \& {Kylafis}(1982)}]{Goldreich1982}
{Goldreich}, P. \& {Kylafis}, N.~D. 1982, \apj, 253, 606

\bibitem[{{Heitsch} {et~al.}(2001){Heitsch}, {Zweibel}, {Mac Low}, {Li}, \&
  {Norman}}]{Heitsch2001}
{Heitsch}, F., {Zweibel}, E.~G., {Mac Low}, M.-M., {Li}, P., \& {Norman}, M.~L.
  2001, \apj, 561, 800

\bibitem[{{Hildebrand}(1988)}]{Hildebrand1988}
{Hildebrand}, R.~H. 1988, \qjras, 29, 327

\bibitem[{{Hildebrand} {et~al.}(1984){Hildebrand}, {Dragovan}, \&
  {Novak}}]{Hildebrand1984}
{Hildebrand}, R.~H., {Dragovan}, M., \& {Novak}, G. 1984, \apjl, 284, L51

\bibitem[{{Hildebrand} {et~al.}(2009){Hildebrand}, {Kirby}, {Dotson}, {Houde},
  \& {Vaillancourt}}]{Hildebrand2009}
{Hildebrand}, R.~H., {Kirby}, L., {Dotson}, J.~L., {Houde}, M., \&
  {Vaillancourt}, J.~E. 2009, \apj, 696, 567

\bibitem[{{Hoang} \& {Lazarian}(2008)}]{Hoang2008}
{Hoang}, T. \& {Lazarian}, A. 2008, \mnras, 388, 117

\bibitem[{{Hofner} {et~al.}(2001){Hofner}, {Wiesemeyer}, \&
  {Henning}}]{Hofner2001}
{Hofner}, P., {Wiesemeyer}, H., \& {Henning}, T. 2001, \apj, 549, 425

\bibitem[{{Houde} {et~al.}(2009){Houde}, {Vaillancourt}, {Hildebrand},
  {Chitsazzadeh}, \& {Kirby}}]{Houde2009}
{Houde}, M., {Vaillancourt}, J.~E., {Hildebrand}, R.~H., {Chitsazzadeh}, S., \&
  {Kirby}, L. 2009, \apj, 706, 1504

\bibitem[{{Hull} {et~al.}(2013){Hull}, {Plambeck}, {Bolatto}, {Bower},
  {Carpenter}, {Crutcher}, {Fiege}, {Franzmann}, {Hakobian}, {Heiles}, {Houde},
  {Hughes}, {Jameson}, {Kwon}, {Lamb}, {Looney}, {Matthews}, {Mundy}, {Pillai},
  {Pound}, {Stephens}, {Tobin}, {Vaillancourt}, {Volgenau}, \&
  {Wright}}]{Hull2013}
{Hull}, C.~L.~H., {Plambeck}, R.~L., {Bolatto}, A.~D., {et~al.} 2013, \apj,
  768, 159

\bibitem[{{Inoue} \& {Fukui}(2013)}]{Inoue2013}
{Inoue}, T. \& {Fukui}, Y. 2013, \apjl, 774, L31

\bibitem[{{Inoue} {et~al.}(2018){Inoue}, {Hennebelle}, {Fukui}, {Matsumoto},
  {Iwasaki}, \& {Inutsuka}}]{Inoue2018}
{Inoue}, T., {Hennebelle}, P., {Fukui}, Y., {et~al.} 2018, \pasj, 70, S53

\bibitem[{Jammalamadaka \& SenGupta(2001)}]{jammalamadaka2001}
Jammalamadaka, S. \& SenGupta, A. 2001, Topics in Circular Statistics, Series
  on multivariate analysis (World Scientific)

\bibitem[{{Johansen} \& {Levin}(2008)}]{Johansen2008}
{Johansen}, A. \& {Levin}, Y. 2008, \aap, 490, 501

\bibitem[{{Ju{\'a}rez} {et~al.}(2017){Ju{\'a}rez}, {Girart},
  {Zamora-Avil{\'e}s}, {Tang}, {Koch}, {Liu}, {Palau}, {Ballesteros-Paredes},
  {Zhang}, \& {Qiu}}]{Juarez2017}
{Ju{\'a}rez}, C., {Girart}, J.~M., {Zamora-Avil{\'e}s}, M., {et~al.} 2017,
  \apj, 844, 44

\bibitem[{{Kataoka} {et~al.}(2016){Kataoka}, {Tsukagoshi}, {Momose}, {Nagai},
  {Muto}, {Dullemond}, {Pohl}, {Fukagawa}, {Shibai}, {Hanawa}, \&
  {Murakawa}}]{Kataoka2016}
{Kataoka}, A., {Tsukagoshi}, T., {Momose}, M., {et~al.} 2016, \apjl, 831, L12

\bibitem[{{Kataoka} {et~al.}(2017){Kataoka}, {Tsukagoshi}, {Pohl}, {Muto},
  {Nagai}, {Stephens}, {Tomisaka}, \& {Momose}}]{Kataoka2017}
{Kataoka}, A., {Tsukagoshi}, T., {Pohl}, A., {et~al.} 2017, \apjl, 844, L5

\bibitem[{{Klassen} {et~al.}(2017){Klassen}, {Pudritz}, \&
  {Kirk}}]{Klassen2017}
{Klassen}, M., {Pudritz}, R.~E., \& {Kirk}, H. 2017, \mnras, 465, 2254

\bibitem[{{Klessen} {et~al.}(2005){Klessen}, {Jappsen}, {Larson}, {Li}, \&
  {Low}}]{Klessen2005}
{Klessen}, R., {Jappsen}, K., {Larson}, R., {Li}, Y., \& {Low}, M.-M.~M. 2005,
  in Astrophysics and Space Science Library, Vol. 327, The Initial Mass
  Function 50 Years Later, ed. E.~{Corbelli}, F.~{Palla}, \& H.~{Zinnecker},
  363--370

\bibitem[{{Koch} {et~al.}(2010){Koch}, {Tang}, \& {Ho}}]{Koch2010}
{Koch}, P.~M., {Tang}, Y.-W., \& {Ho}, P.~T.~P. 2010, \apj, 721, 815

\bibitem[{{Lai} {et~al.}(2003){Lai}, {Girart}, \& {Crutcher}}]{Lai2003}
{Lai}, S.-P., {Girart}, J.~M., \& {Crutcher}, R.~M. 2003, \apj, 598, 392

\bibitem[{{Lazarian}(2000)}]{Lazarian2000}
{Lazarian}, A. 2000, in Astronomical Society of the Pacific Conference Series,
  Vol. 215, Cosmic Evolution and Galaxy Formation: Structure, Interactions, and
  Feedback, ed. J.~{Franco}, L.~{Terlevich}, O.~{L{\'o}pez-Cruz}, \&
  I.~{Aretxaga}, 69

\bibitem[{{Lazarian} \& {Hoang}(2007)}]{Lazarian2007}
{Lazarian}, A. \& {Hoang}, T. 2007, \apjl, 669, L77

\bibitem[{{Lazarian} \& {Hoang}(2008)}]{Lazarian2008}
{Lazarian}, A. \& {Hoang}, T. 2008, \apjl, 676, L25

\bibitem[{{Linz} {et~al.}(2005){Linz}, {Stecklum}, {Henning}, {Hofner}, \&
  {Brandl}}]{Linz2005}
{Linz}, H., {Stecklum}, B., {Henning}, T., {Hofner}, P., \& {Brandl}, B. 2005,
  \aap, 429, 903

\bibitem[{{Liu} {et~al.}(2018){Liu}, {Kim}, {Liu}, {Juvela}, {Zhang}, {Wu},
  {Li}, {Parsons}, {Soam}, {Goldsmith}, {Su}, {Tatematsu}, {Qin}, {Garay},
  {Hirota}, {Wouterloot}, {Chen}, {Evans}, {Graves}, {Kang}, {Li}, {Mardones},
  {Rawlings}, {Ren}, \& {Wang}}]{Liu2018}
{Liu}, T., {Kim}, K.-T., {Liu}, S.-Y., {et~al.} 2018, \apjl, 869, L5

\bibitem[{{Liu} {et~al.}(2017){Liu}, {Lacy}, {Li}, {Wang}, {Qin}, {Zhang},
  {Kim}, {Garay}, {Wu}, {Mardones}, {Zhu}, {Tatematsu}, {Hirota}, {Ren}, {Liu},
  {Chen}, {Su}, \& {Li}}]{Liu2017}
{Liu}, T., {Lacy}, J., {Li}, P.~S., {et~al.} 2017, \apj, 849, 25

\bibitem[{{Liu} {et~al.}(2011){Liu}, {Wu}, {Liu}, {Qin}, {Su}, {Chen}, \&
  {Ren}}]{Liu2011}
{Liu}, T., {Wu}, Y., {Liu}, S.-Y., {et~al.} 2011, \apj, 730, 102

\bibitem[{{Machida} {et~al.}(2014){Machida}, {Inutsuka}, \&
  {Matsumoto}}]{Machida2014}
{Machida}, M.~N., {Inutsuka}, S.-i., \& {Matsumoto}, T. 2014, \mnras, 438, 2278

\bibitem[{{Matsushita} {et~al.}(2018){Matsushita}, {Sakurai}, {Hosokawa}, \&
  {Machida}}]{Matsushita2018}
{Matsushita}, Y., {Sakurai}, Y., {Hosokawa}, T., \& {Machida}, M.~N. 2018,
  \mnras, 475, 391

\bibitem[{{Maury} {et~al.}(2018){Maury}, {Girart}, {Zhang}, {Hennebelle},
  {Keto}, {Rao}, {Lai}, {Ohashi}, \& {Galametz}}]{Maury2018}
{Maury}, A.~J., {Girart}, J.~M., {Zhang}, Q., {et~al.} 2018, \mnras, 477, 2760

\bibitem[{{McKee} \& {Ostriker}(2007)}]{McKee2007}
{McKee}, C.~F. \& {Ostriker}, E.~C. 2007, \araa, 45, 565

\bibitem[{{McKee} \& {Tan}(2003)}]{McKeeTan2002}
{McKee}, C.~F. \& {Tan}, J.~C. 2003, \apj, 585, 850

\bibitem[{Mouschovias(1991)}]{Mouschovias1991}
Mouschovias, T.~C. 1991, Cosmic Magnetism and the Basic Physics of the Early
  Stages of Star Formation, ed. C.~J. Lada \& N.~D. Kylafis (Dordrecht:
  Springer Netherlands), 61--122

\bibitem[{{Mouschovias} \& {Paleologou}(1979)}]{Mouschovias1979}
{Mouschovias}, T.~C. \& {Paleologou}, E.~V. 1979, \apj, 230, 204

\bibitem[{{Mouschovias} {et~al.}(2006){Mouschovias}, {Tassis}, \&
  {Kunz}}]{Mouschovias2006}
{Mouschovias}, T.~C., {Tassis}, K., \& {Kunz}, M.~W. 2006, \apj, 646, 1043

\bibitem[{{Myers} {et~al.}(2013){Myers}, {McKee}, {Cunningham}, {Klein}, \&
  {Krumholz}}]{Myers2013}
{Myers}, A.~T., {McKee}, C.~F., {Cunningham}, A.~J., {Klein}, R.~I., \&
  {Krumholz}, M.~R. 2013, \apj, 766, 97

\bibitem[{{Nakano} \& {Nakamura}(1978)}]{Nakano1978}
{Nakano}, T. \& {Nakamura}, T. 1978, \pasj, 30, 671

\bibitem[{{Ossenkopf} \& {Henning}(1994)}]{Ossenkopf1994}
{Ossenkopf}, V. \& {Henning}, T. 1994, \aap, 291, 943

\bibitem[{{Ostriker} {et~al.}(2001){Ostriker}, {Stone}, \&
  {Gammie}}]{Ostriker2001}
{Ostriker}, E.~C., {Stone}, J.~M., \& {Gammie}, C.~F. 2001, \apj, 546, 980

\bibitem[{{Padoan} {et~al.}(2001){Padoan}, {Goodman}, {Draine}, {Juvela},
  {Nordlund}, \& {R{\"o}gnvaldsson}}]{Padoan2001}
{Padoan}, P., {Goodman}, A., {Draine}, B.~T., {et~al.} 2001, \apj, 559, 1005

\bibitem[{{Padoan} \& {Nordlund}(2002)}]{Padoan2002}
{Padoan}, P. \& {Nordlund}, {\AA}. 2002, \apj, 576, 870

\bibitem[{{Persi} {et~al.}(2003){Persi}, {Tapia}, {Roth}, {Marenzi}, {Testi},
  \& {Vanzi}}]{Persi2003}
{Persi}, P., {Tapia}, M., {Roth}, M., {et~al.} 2003, \aap, 397, 227

\bibitem[{{Peters} {et~al.}(2014){Peters}, {Schleicher}, {Smith}, {Schmidt}, \&
  {Klessen}}]{Peters2014}
{Peters}, T., {Schleicher}, D.~R.~G., {Smith}, R.~J., {Schmidt}, W., \&
  {Klessen}, R.~S. 2014, \mnras, 442, 3112

\bibitem[{Pewsey {et~al.}(2013)Pewsey, Neuh{\"a}user, \& Ruxton}]{pewsey2013}
Pewsey, A., Neuh{\"a}user, M., \& Ruxton, G. 2013, Circular Statistics in R,
  EBL-Schweitzer (OUP Oxford)

\bibitem[{{Planck Collaboration} {et~al.}(2016){Planck Collaboration}, {Ade},
  {Aghanim}, {Alves}, {Arnaud}, {Arzoumanian}, {Ashdown}, {Aumont},
  {Baccigalupi}, {Banday}, {Barreiro}, {Bartolo}, {Battaner}, {Benabed},
  {Beno{\^i}t}, {Benoit-L{\'e}vy}, {Bernard}, {Bersanelli}, {Bielewicz},
  {Bock}, {Bonavera}, {Bond}, {Borrill}, {Bouchet}, {Boulanger}, {Bracco},
  {Burigana}, {Calabrese}, {Cardoso}, {Catalano}, {Chiang}, {Christensen},
  {Colombo}, {Combet}, {Couchot}, {Crill}, {Curto}, {Cuttaia}, {Danese},
  {Davies}, {Davis}, {de Bernardis}, {de Rosa}, {de Zotti}, {Delabrouille},
  {Dickinson}, {Diego}, {Dole}, {Donzelli}, {Dor{\'e}}, {Douspis}, {Ducout},
  {Dupac}, {Efstathiou}, {Elsner}, {En{\ss}lin}, {Eriksen}, {Falceta-Gon{\c
  c}alves}, {Falgarone}, {Ferri{\`e}re}, {Finelli}, {Forni}, {Frailis},
  {Fraisse}, {Franceschi}, {Frejsel}, {Galeotta}, {Galli}, {Ganga}, {Ghosh},
  {Giard}, {Gjerl{\o}w}, {Gonz{\'a}lez-Nuevo}, {G{\'o}rski}, {Gregorio},
  {Gruppuso}, {Gudmundsson}, {Guillet}, {Harrison}, {Helou}, {Hennebelle},
  {Henrot-Versill{\'e}}, {Hern{\'a}ndez-Monteagudo}, {Herranz}, {Hildebrandt},
  {Hivon}, {Holmes}, {Hornstrup}, {Huffenberger}, {Hurier}, {Jaffe}, {Jaffe},
  {Jones}, {Juvela}, {Keih{\"a}nen}, {Keskitalo}, {Kisner}, {Knoche}, {Kunz},
  {Kurki-Suonio}, {Lagache}, {Lamarre}, {Lasenby}, {Lattanzi}, {Lawrence},
  {Leonardi}, {Levrier}, {Liguori}, {Lilje}, {Linden-V{\o}rnle},
  {L{\'o}pez-Caniego}, {Lubin}, {Mac{\'{\i}}as-P{\'e}rez}, {Maino},
  {Mandolesi}, {Mangilli}, {Maris}, {Martin}, {Mart{\'{\i}}nez-Gonz{\'a}lez},
  {Masi}, {Matarrese}, {Melchiorri}, {Mendes}, {Mennella}, {Migliaccio},
  {Miville-Desch{\^e}nes}, {Moneti}, {Montier}, {Morgante}, {Mortlock},
  {Munshi}, {Murphy}, {Naselsky}, {Nati}, {Netterfield}, {Noviello}, {Novikov},
  {Novikov}, {Oppermann}, {Oxborrow}, {Pagano}, {Pajot}, {Paladini},
  {Paoletti}, {Pasian}, {Perotto}, {Pettorino}, {Piacentini}, {Piat},
  {Pierpaoli}, {Pietrobon}, {Plaszczynski}, {Pointecouteau}, {Polenta},
  {Ponthieu}, {Pratt}, {Prunet}, {Puget}, {Rachen}, {Reinecke}, {Remazeilles},
  {Renault}, {Renzi}, {Ristorcelli}, {Rocha}, {Rossetti}, {Roudier},
  {Rubi{\~n}o-Mart{\'{\i}}n}, {Rusholme}, {Sandri}, {Santos}, {Savelainen},
  {Savini}, {Scott}, {Soler}, {Stolyarov}, {Sudiwala}, {Sutton}, {Suur-Uski},
  {Sygnet}, {Tauber}, {Terenzi}, {Toffolatti}, {Tomasi}, {Tristram}, {Tucci},
  {Umana}, {Valenziano}, {Valiviita}, {Van Tent}, {Vielva}, {Villa}, {Wade},
  {Wandelt}, {Wehus}, {Ysard}, {Yvon}, \& {Zonca}}]{Planck_coll2016}
{Planck Collaboration}, {Ade}, P.~A.~R., {Aghanim}, N., {et~al.} 2016, \aap,
  586, A138

\bibitem[{{Qin} {et~al.}(2010){Qin}, {Wu}, {Huang}, {Zhao}, {Li}, {Wang}, \&
  {Chen}}]{Qin2010}
{Qin}, S.-L., {Wu}, Y., {Huang}, M., {et~al.} 2010, \apj, 711, 399

\bibitem[{{Qiu} {et~al.}(2014){Qiu}, {Zhang}, {Menten}, {Liu}, {Tang}, \&
  {Girart}}]{Qiu2014}
{Qiu}, K., {Zhang}, Q., {Menten}, K.~M., {et~al.} 2014, \apjl, 794, L18

\bibitem[{{Sanna} {et~al.}(2009){Sanna}, {Reid}, {Moscadelli}, {Dame},
  {Menten}, {Brunthaler}, {Zheng}, \& {Xu}}]{Sanna2009}
{Sanna}, A., {Reid}, M.~J., {Moscadelli}, L., {et~al.} 2009, \apj, 706, 464

\bibitem[{{Seifried} {et~al.}(2012){Seifried}, {Banerjee}, {Pudritz}, \&
  {Klessen}}]{Seifried2012}
{Seifried}, D., {Banerjee}, R., {Pudritz}, R.~E., \& {Klessen}, R.~S. 2012,
  \mnras, 423, L40

\bibitem[{{Stepanovs} {et~al.}(2014){Stepanovs}, {Fendt}, \&
  {Sheikhnezami}}]{Stepanovs2014}
{Stepanovs}, D., {Fendt}, C., \& {Sheikhnezami}, S. 2014, \apj, 796, 29

\bibitem[{{Susa} {et~al.}(2015){Susa}, {Doi}, \& {Omukai}}]{Susa2015}
{Susa}, H., {Doi}, K., \& {Omukai}, K. 2015, \apj, 801, 13

\bibitem[{{Tan} {et~al.}(2013){Tan}, {Kong}, {Butler}, {Caselli}, \&
  {Fontani}}]{Tan2013}
{Tan}, J.~C., {Kong}, S., {Butler}, M.~J., {Caselli}, P., \& {Fontani}, F.
  2013, \apj, 779, 96

\bibitem[{{Tassis} {et~al.}(2014){Tassis}, {Willacy}, {Yorke}, \&
  {Turner}}]{Tassis2014}
{Tassis}, K., {Willacy}, K., {Yorke}, H.~W., \& {Turner}, N.~J. 2014, \mnras,
  445, L56

\bibitem[{{Testi} {et~al.}(1998){Testi}, {Felli}, {Persi}, \&
  {Roth}}]{Testi1998}
{Testi}, L., {Felli}, M., {Persi}, P., \& {Roth}, M. 1998, \aap, 329, 233

\bibitem[{{Testi} {et~al.}(2000){Testi}, {Hofner}, {Kurtz}, \&
  {Rupen}}]{Testi2000}
{Testi}, L., {Hofner}, P., {Kurtz}, S., \& {Rupen}, M. 2000, \aap, 359, L5

\bibitem[{{Thompson} {et~al.}(2006){Thompson}, {Hatchell}, {Walsh},
  {MacDonald}, \& {Millar}}]{Thompson2006}
{Thompson}, M.~A., {Hatchell}, J., {Walsh}, A.~J., {MacDonald}, G.~H., \&
  {Millar}, T.~J. 2006, \aap, 453, 1003

\bibitem[{{Troland} \& {Crutcher}(2008)}]{Troland2008}
{Troland}, T.~H. \& {Crutcher}, R.~M. 2008, \apj, 680, 457

\bibitem[{{Vaidya} {et~al.}(2013){Vaidya}, {Hartquist}, \&
  {Falle}}]{Vaidya2013}
{Vaidya}, B., {Hartquist}, T.~W., \& {Falle}, S.~A.~E.~G. 2013, \mnras, 433,
  1258

\bibitem[{{V{\'a}zquez-Semadeni} {et~al.}(2011){V{\'a}zquez-Semadeni},
  {Banerjee}, {G{\'o}mez}, {Hennebelle}, {Duffin}, \&
  {Klessen}}]{Vazquez-Semadeni2011}
{V{\'a}zquez-Semadeni}, E., {Banerjee}, R., {G{\'o}mez}, G.~C., {et~al.} 2011,
  \mnras, 414, 2511

\bibitem[{{Vlemmings}(2008)}]{Vlemmings2008}
{Vlemmings}, W.~H.~T. 2008, \aap, 484, 773

\bibitem[{{Wang} {et~al.}(2014){Wang}, {Zhang}, {Testi}, {van der Tak}, {Wu},
  {Zhang}, {Pillai}, {Wyrowski}, {Carey}, {Ragan}, \& {Henning}}]{Wang2014}
{Wang}, K., {Zhang}, Q., {Testi}, L., {et~al.} 2014, \mnras, 439, 3275

\bibitem[{{Wardle} \& {Kronberg}(1974)}]{Wardle&Kronberg1974}
{Wardle}, J.~F.~C. \& {Kronberg}, P.~P. 1974, \apj, 194, 249

\bibitem[{{Williams} {et~al.}(2004){Williams}, {Fuller}, \&
  {Sridharan}}]{Williams2004}
{Williams}, S.~J., {Fuller}, G.~A., \& {Sridharan}, T.~K. 2004, \aap, 417, 115

\bibitem[{{Zhang} {et~al.}(2014){Zhang}, {Qiu}, {Girart}, {Liu}, {Tang},
  {Koch}, {Li}, {Keto}, {Ho}, {Rao}, {Lai}, {Ching}, {Frau}, {Chen}, {Li},
  {Padovani}, {Bontemps}, {Csengeri}, \& {Ju{\'a}rez}}]{Zhang2014}
{Zhang}, Q., {Qiu}, K., {Girart}, J.~M., {et~al.} 2014, \apj, 792, 116

\bibitem[{{Zhao} {et~al.}(2016){Zhao}, {Caselli}, {Li}, {Krasnopolsky},
  {Shang}, \& {Nakamura}}]{Zhao2016}
{Zhao}, B., {Caselli}, P., {Li}, Z.-Y., {et~al.} 2016, \mnras, 460, 2050

\end{thebibliography}

\end{document}